\documentclass[aps,prd,twocolumn,showpacs,amsmath,amssymb]{revtex4-1}

\usepackage[dvips]{graphicx}

\begin{document}

\title
{
  Intrinsic scatter of the luminosity relation,
  redshift distribution of the standard candles,
  and the constraining capability
}
\author{Shi Qi}
\email{qishi11@gmail.com}
\affiliation
{
  Purple Mountain Observatory,
  Chinese Academy of Sciences,
  Nanjing 210008,
  China
}
\affiliation
{
  Key Laboratory of Dark Matter and Space Astronomy,
  Chinese Academy of Sciences,
  Nanjing 210008,
  China
}
\affiliation
{
  Joint Center for Particle, Nuclear Physics and Cosmology,
  Nanjing University---Purple Mountain Observatory,
  Nanjing 210093,
  China
}
\affiliation
{
  Kavli Institute for Theoretical Physics China,
  Chinese Academy of Sciences,
  Beijing 100190,
  China
}

% \date{\today}

\begin{abstract}
  Standard candles are one of the most important tools to study the
  universe.
  In this paper, the constraints of standards candles on the
  cosmological parameters are estimated for different cases.
  The dependence of the constraints on the intrinsic scatter of the
  luminosity relation and the redshift distribution of the standard
  candles is specifically investigated.
  The results, especially for the constraints on the components of the
  universe, clearly show that constraints from standard candles at
  different redshifts have different degeneracy orientations,
  thus standard candles with a wide redshift distribution can self
  break the degeneracy and improve the constraints significantly.
  As a result of this,
  even with the current level of tightness of known luminosity
  relations,
  gamma-ray bursts (GRBs) can give comparable tightness of constraint
  with type Ia supernovae (SNe Ia) on the components of the universe
  as long as the redshifts of the GRBs are diversifying enough.
  However, for a substantial constraint on the dark energy EOS,
  tighter luminosity relations for GRBs are needed,
  since the constraints on the dark energy from standard candles at
  high redshifts are very weak and are thus less helpful in the
  degeneracy breaking.
\end{abstract}

\pacs{}

\keywords{}

\maketitle

Standard candles are one of the most important tools to study the
universe.
Type Ia supernovae (SNe Ia) are currently the maturest standard
candles on cosmological scales, studies on which lead to the discovery
of cosmic acceleration~\cite{Riess:1998cb,Perlmutter:1998np}.
Gamma-ray bursts (GRBs) have also attracted much attention as standard
candles (see e.g.~\cite{Qi:2011tr} and references therein).
GRBs cover much wider redshift range than SNe Ia,
but, on the other hand, have larger intrinsic scatters in their
luminosity relations,
which makes them less ideal as standard candles than SNe Ia.
In this paper, the constraints of standards candles on the
cosmological parameters are estimated for different cases.
The dependence of the constraints on the intrinsic scatter of the
luminosity relation and the redshift distribution of the standard
candles is specifically investigated.
The investigation is done by keeping in mind the current development
of the GRBs as standard candles (see e.g.~\cite{Qi:2011tr}).

In~\cite{Qi:2015uxa}, a general procedure for estimating constraints
of standard candles on cosmological parameters using mock data was
discussed and,
as a result, analytical formulae for the marginal likelihood of the
cosmological parameters were derived.
Consider a general luminosity relation of the form
\begin{equation}
  \label{eq:lr}
  y = c_0 + \sum_i c_i x_i + \varepsilon
  ,
\end{equation}
where $x_i$s are some luminosity indicators which can be directly
measured from observation,
$\varepsilon$ is a random variable accounting for the intrinsic
scatter $\sigma_{\mathrm{int}, 0}$ of the relation,
and $y$ has the form of
\begin{equation}
  \label{eq:y_dl}
  y = \log
  \left(
    4 \pi d_L^2 \mathcal{F}
  \right)
  ,
\end{equation}
where $d_L$ is the luminosity distance and $\mathcal{F}$ may be any
physical quantity that can be directly measured from observation.
Define
\begin{equation}
  \label{eq:lztheta_dl}
  l(z, \theta, \theta_0)
  =
  2 \log
  \frac
  {
    d_L(z, \theta)
  }
  {
    d_L(z, \theta_0)
  }
\end{equation}
and use $l_i$ as the abbreviation for $l(z_i, \theta, \theta_0)$.
Ignoring the measurement uncertainties, we have the marginal
likelihood of the cosmological parameters
$\theta$
\begin{equation}
  \label{eq:Ltheta}
  \mathcal{L}(\theta)
  \propto
  \left(
    \sigma_{\mathrm{int}, 0}^2 + \sigma_l^2
  \right)^{- \frac{N-p}{2}}
  ,
\end{equation}
where $p$ is the number of the calibration parameters which include
the coefficients $c$ and the intrinsic scatter
$\sigma_{\mathrm{int}}$.
See~\cite{Qi:2015uxa} for more details.

From Eqs.~(\ref{eq:lztheta_dl}) and (\ref{eq:Ltheta}), we can see
that, to estimate the constraining capability of a sample of standard
candles on the cosmological parameters (without considering the
measurement uncertainties), we only need to input the information
\begin{enumerate}
\item about the luminosity relation: its intrinsic scatter and the
  number of luminosity indicators involved;
\item about the sample: the number of the standard candles and their
  redshifts.
\end{enumerate}
No further detailed information is needed.
This much simplifies the procedure and makes things transparent and
clear.
For example, one can immediately tell from Eq.~(\ref{eq:Ltheta}) that
the smaller the intrinsic scatter $\sigma_{\mathrm{int}, 0}$ is and/or
the larger the sample size $N$ is, the more sensitive the marginal
likelihood $\mathcal{L}(\theta)$ is to the variation of the
cosmological parameters $\theta$, which means that the constraint is
tighter.
The Hubble constant only contributes to
$l(z, \theta, \theta_0)$ a constant
that is same for all the standard candles
and has no effect on $\sigma_l$,
so $\mathcal{L}(\theta)$ is independent of the Hubble constant and we
cannot directly constrain it in this way.

\begin{figure*}[tbp]
  \centering
  \includegraphics[width = 0.19 \textwidth]{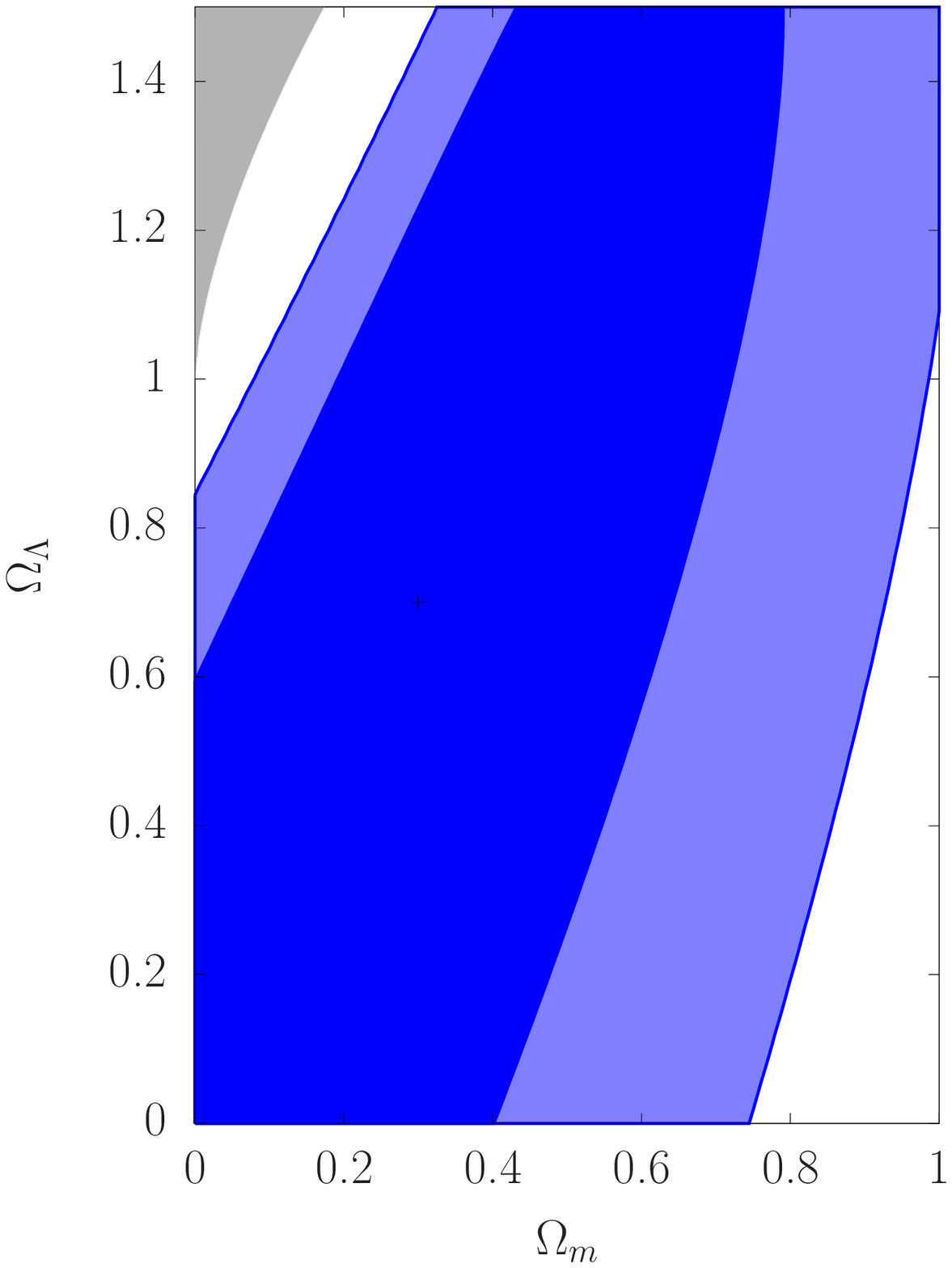}
  \includegraphics[width = 0.19 \textwidth]{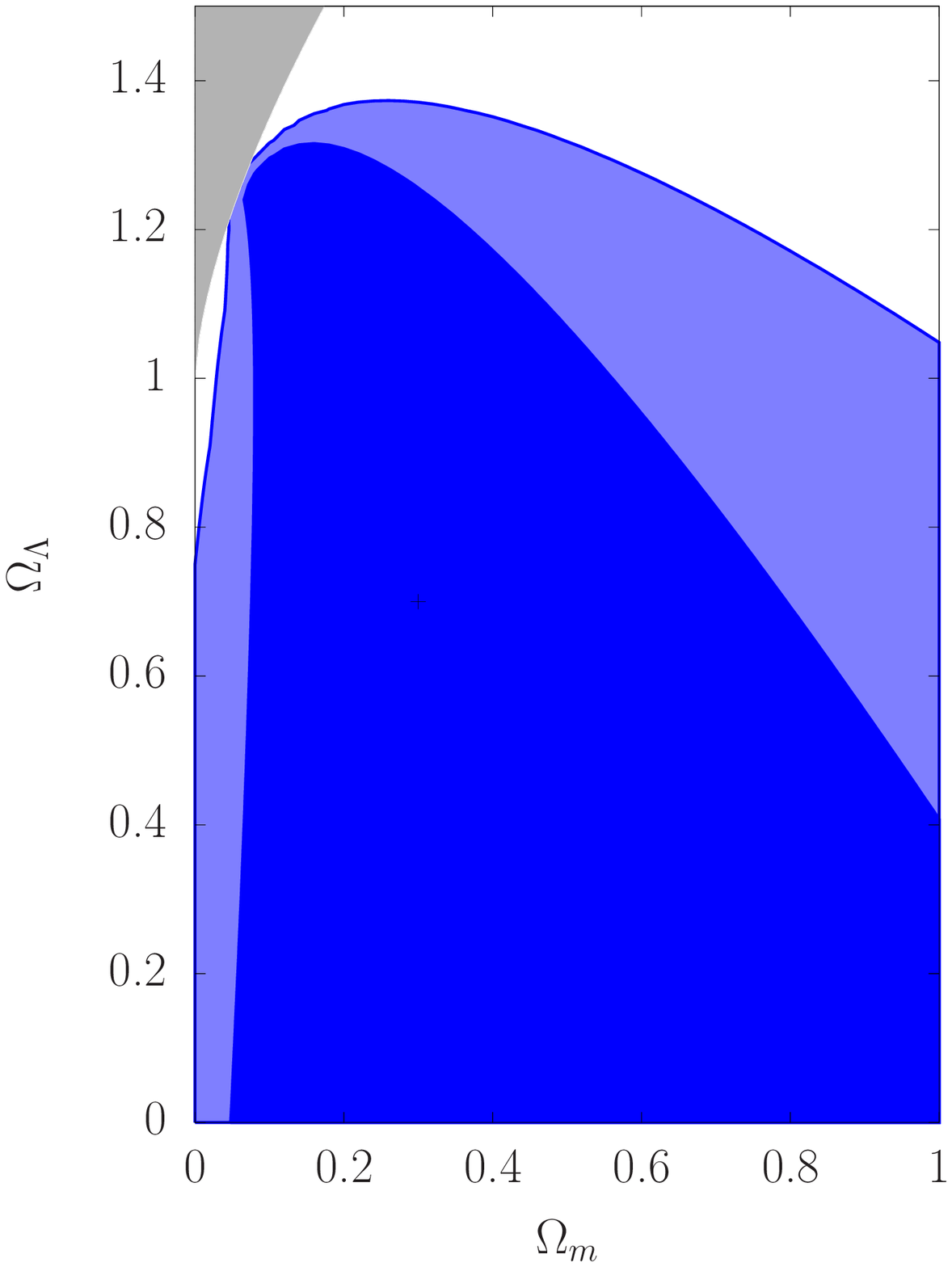}
  \includegraphics[width = 0.19 \textwidth]{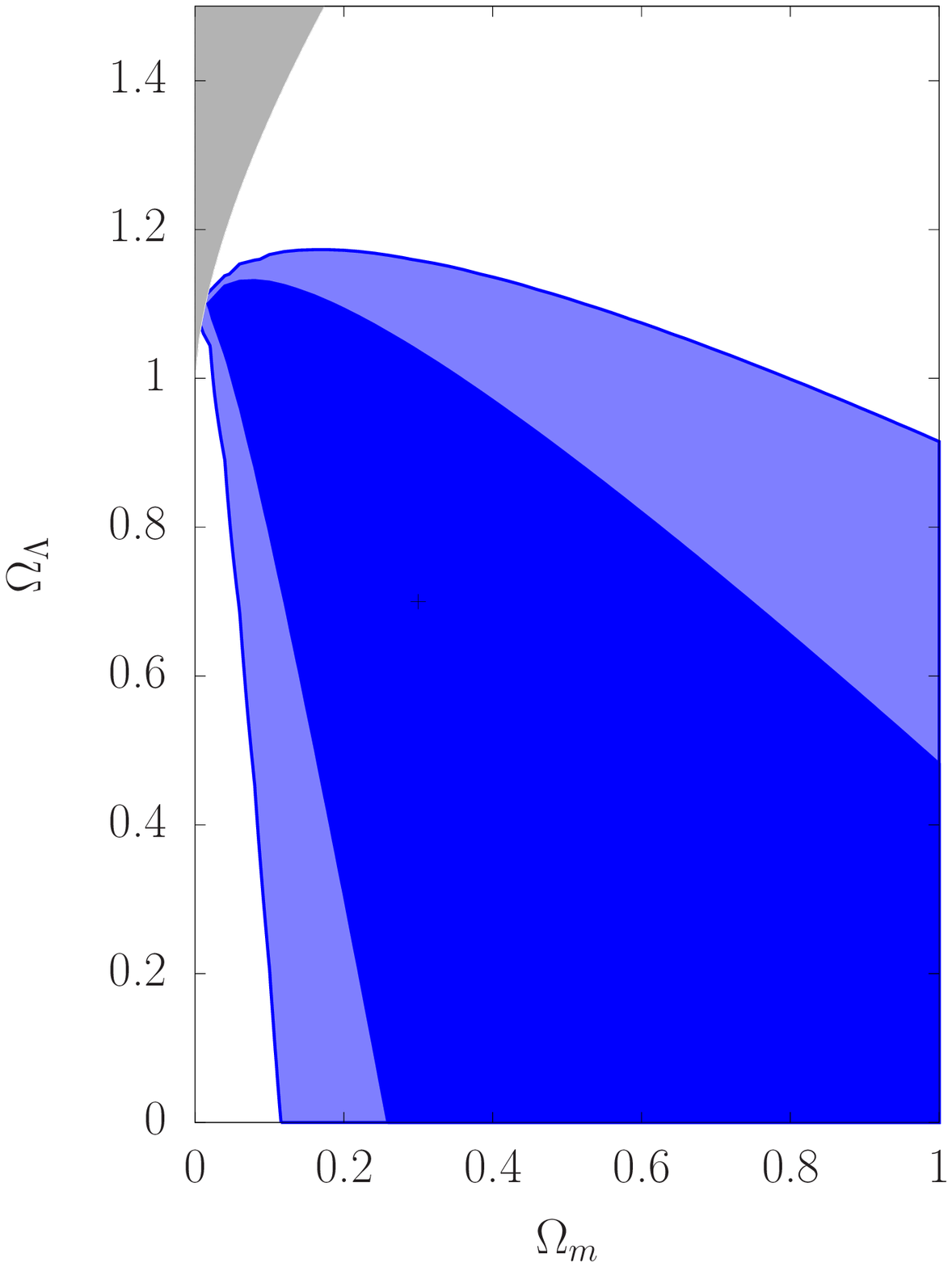}
  \includegraphics[width = 0.19 \textwidth]{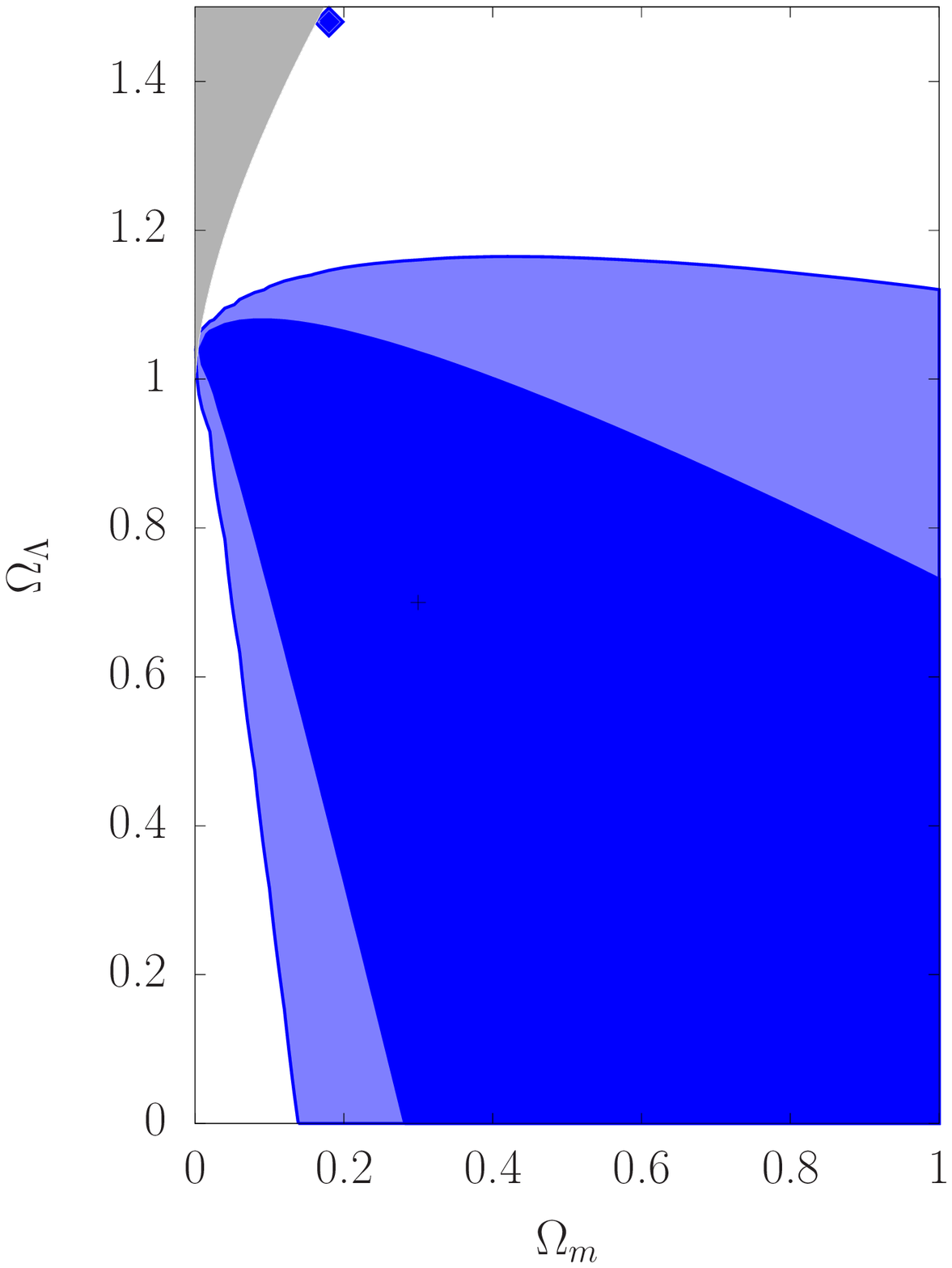}
  \includegraphics[width = 0.19 \textwidth]{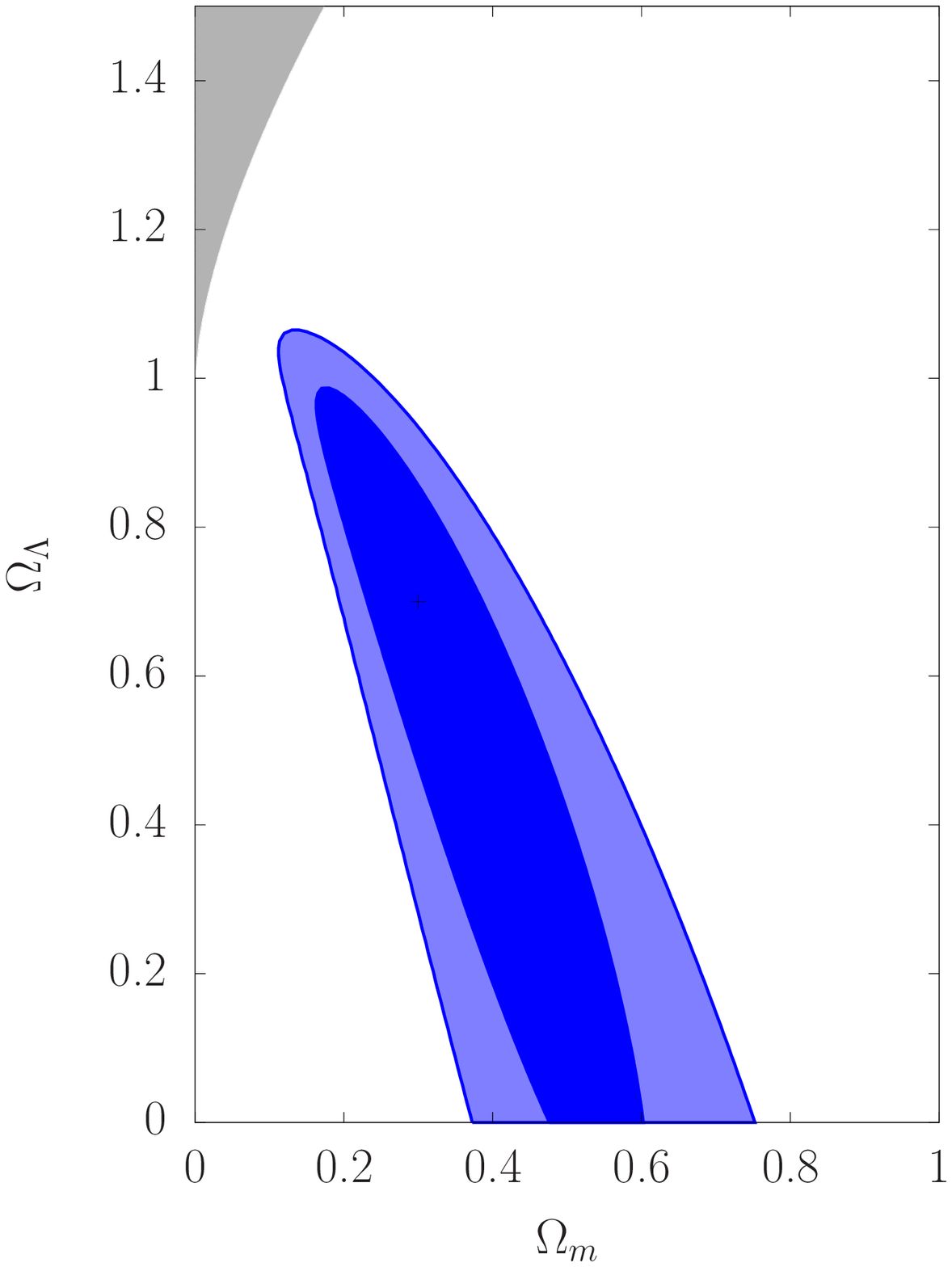}
  \\
  \includegraphics[width = 0.19 \textwidth]{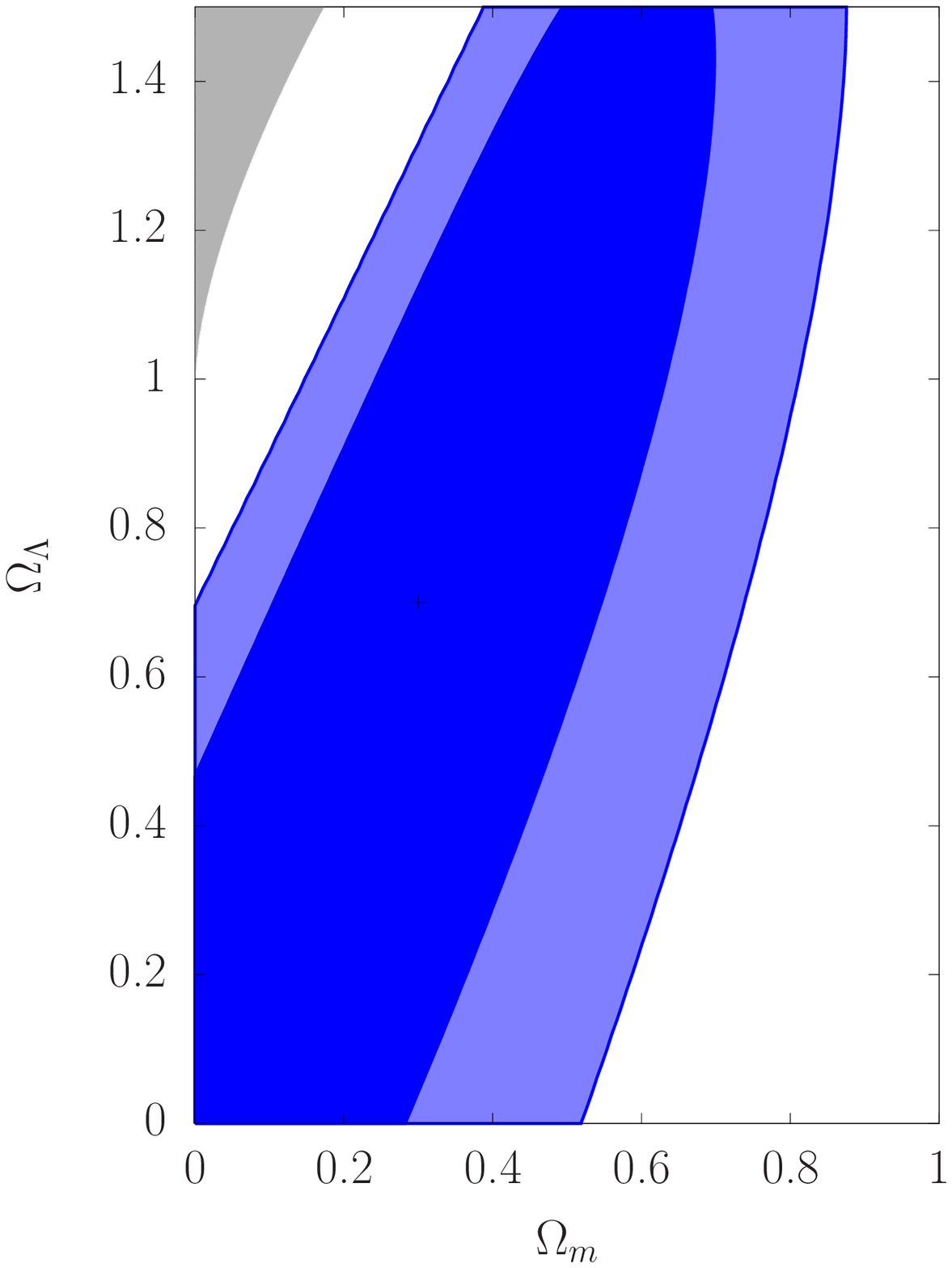}
  \includegraphics[width = 0.19 \textwidth]{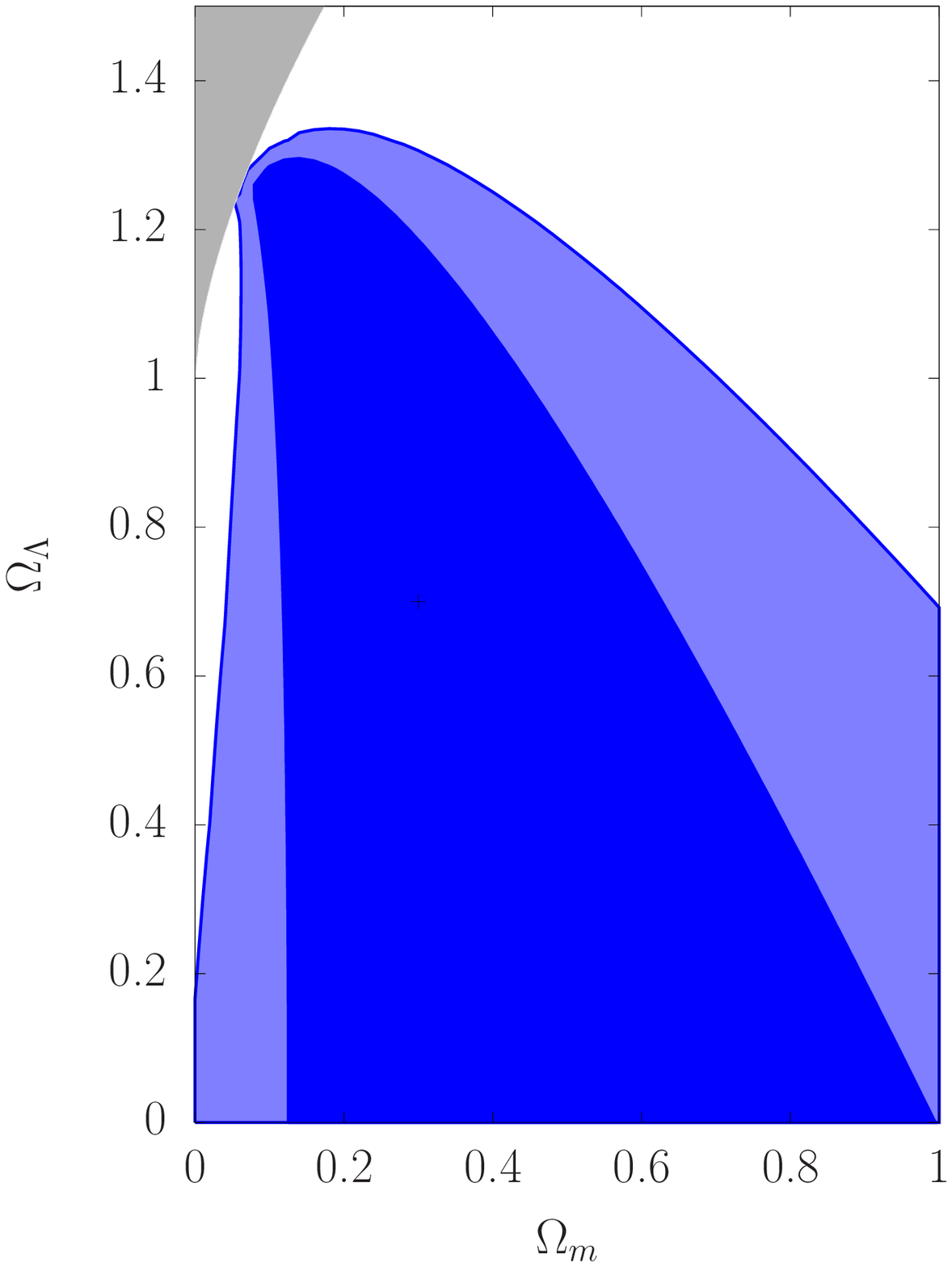}
  \includegraphics[width = 0.19 \textwidth]{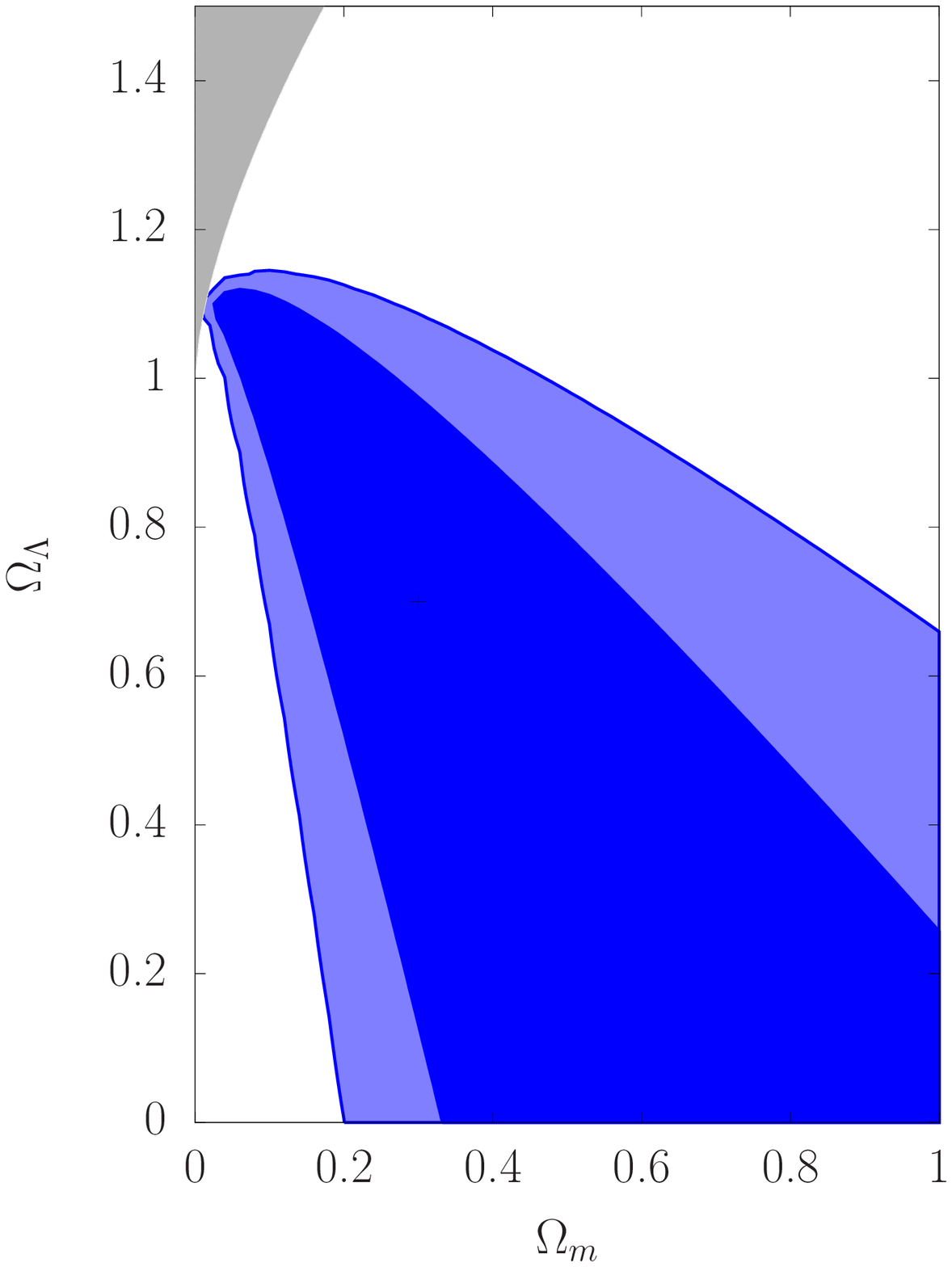}
  \includegraphics[width = 0.19 \textwidth]{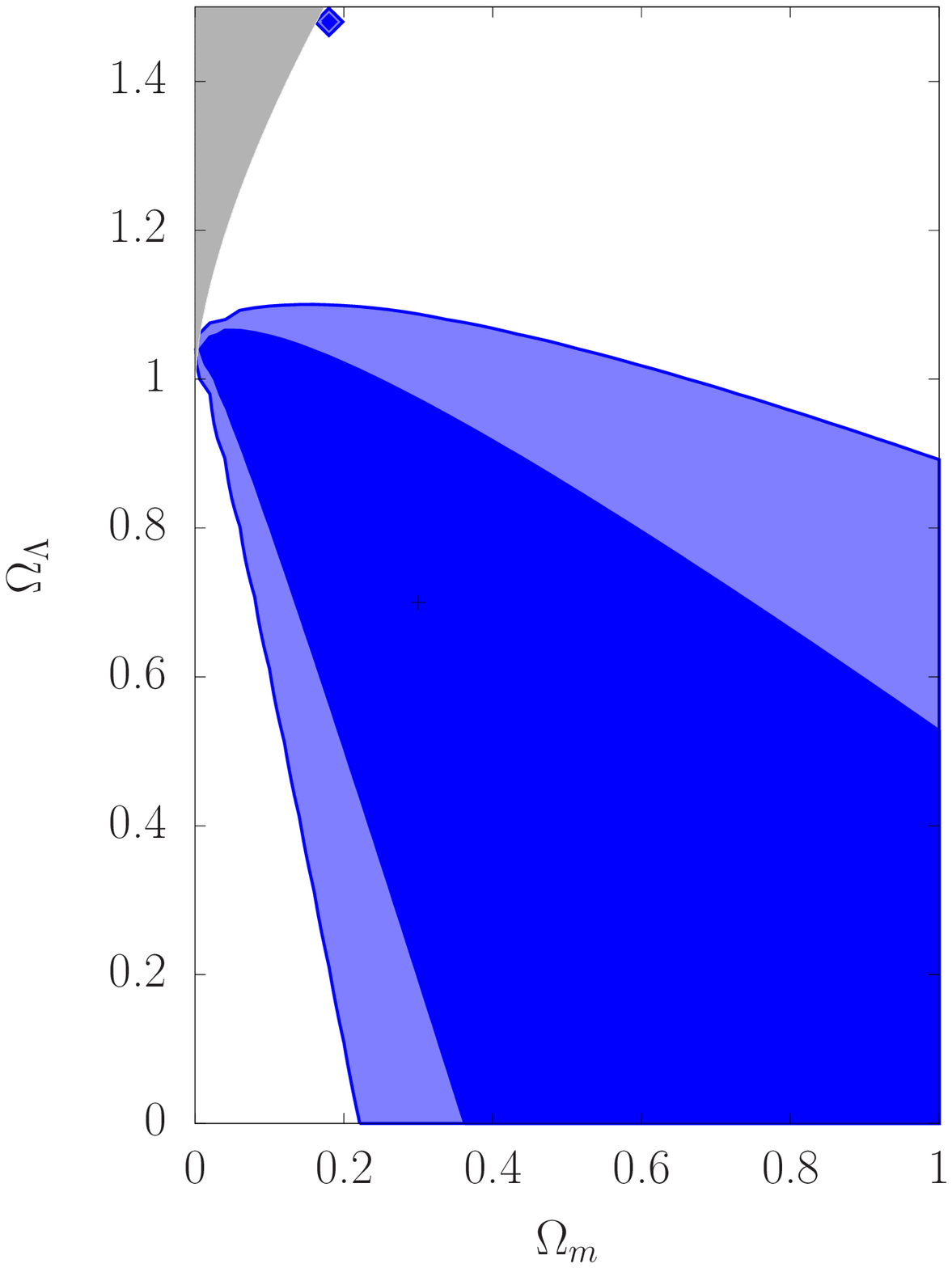}
  \includegraphics[width = 0.19 \textwidth]{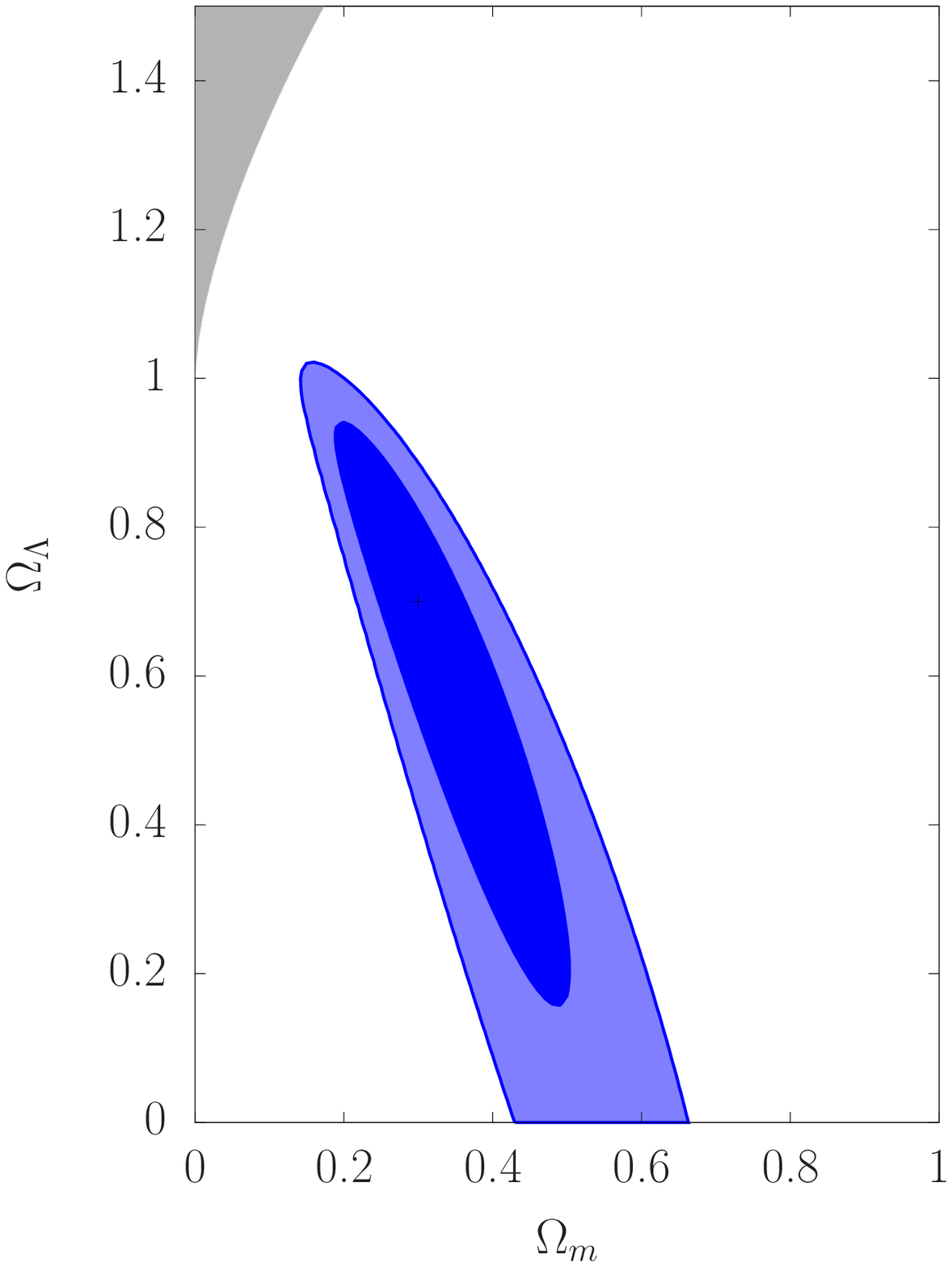}
  \\
  \includegraphics[width = 0.19 \textwidth]{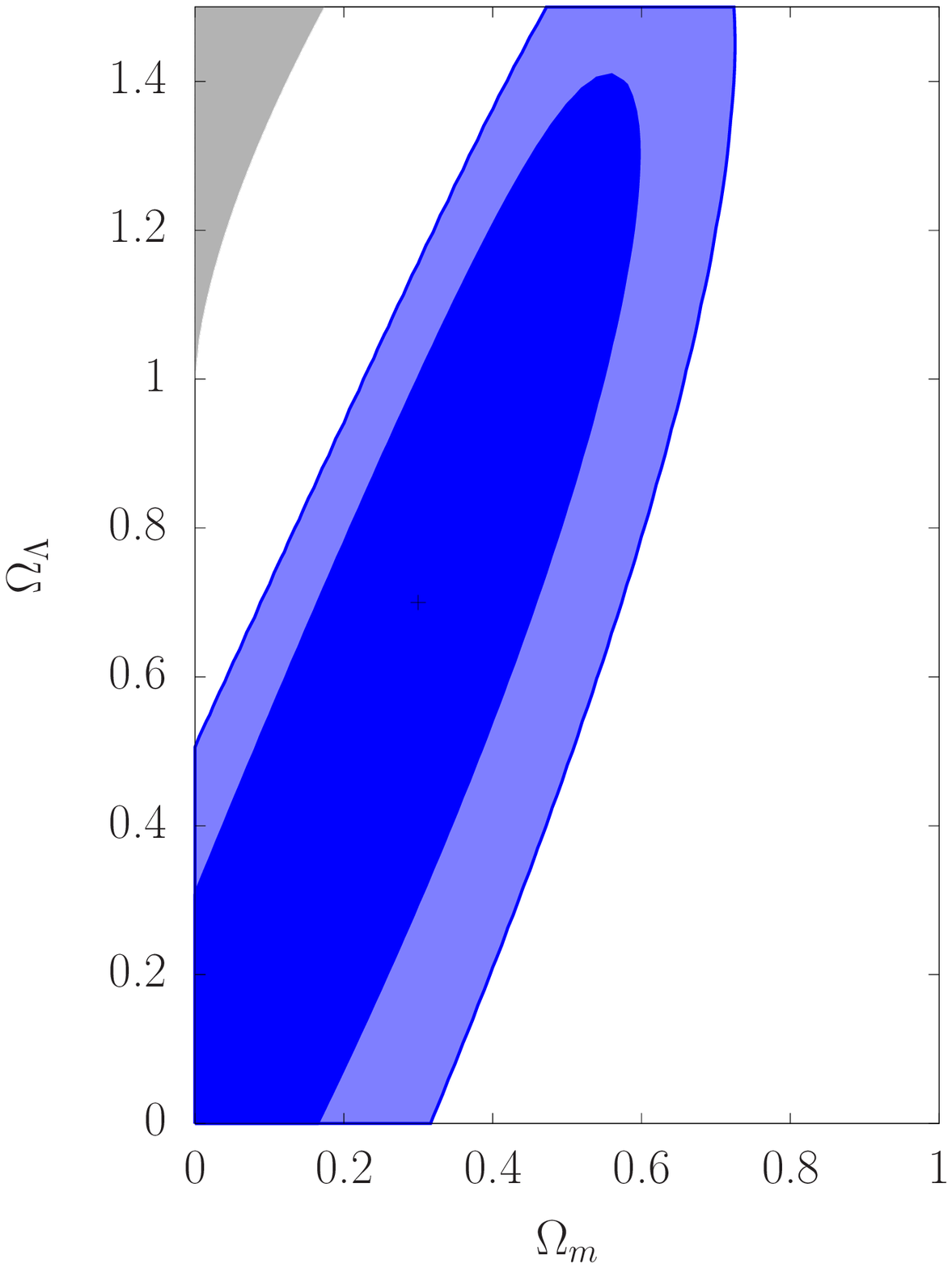}
  \includegraphics[width = 0.19 \textwidth]{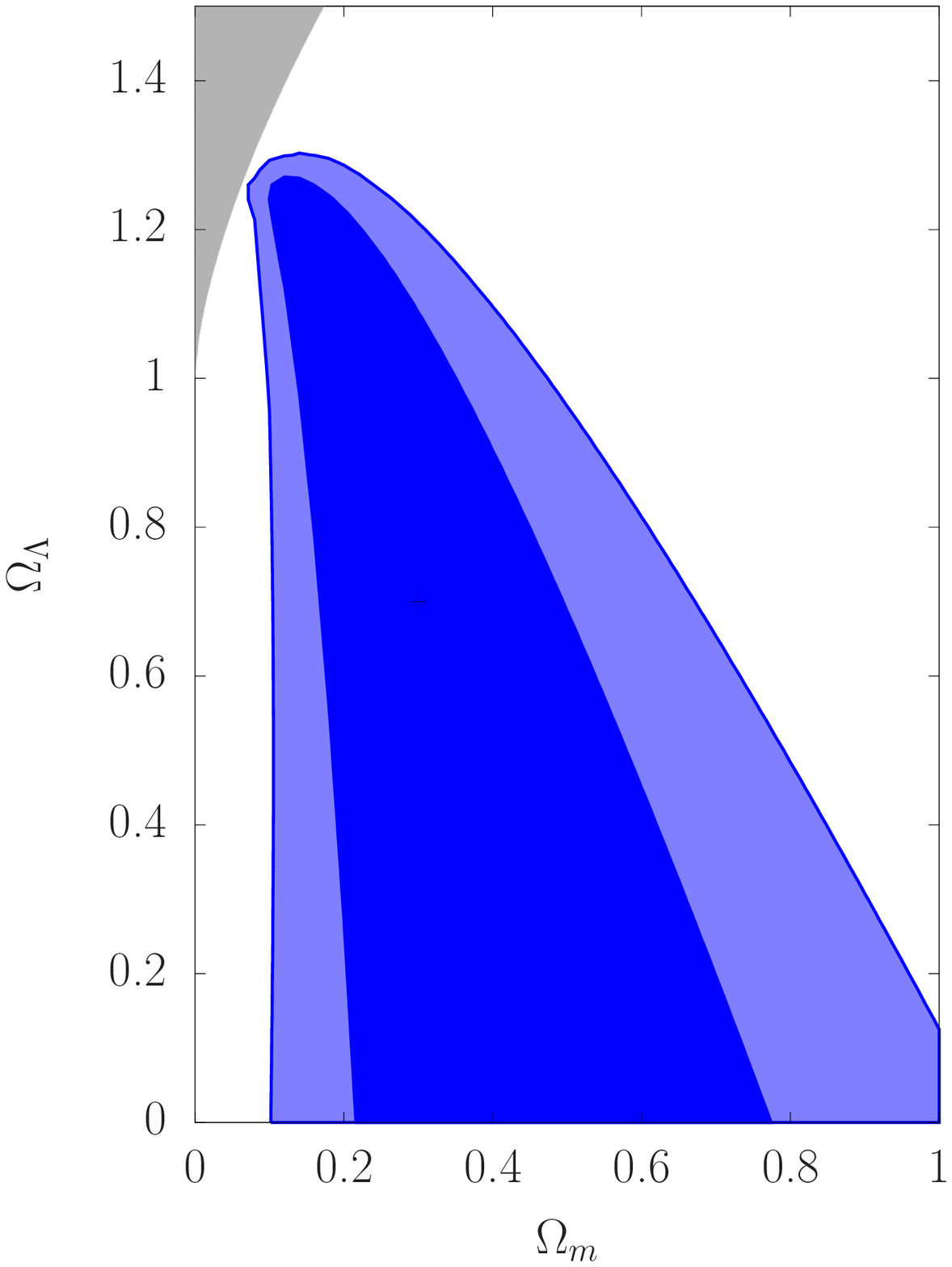}
  \includegraphics[width = 0.19 \textwidth]{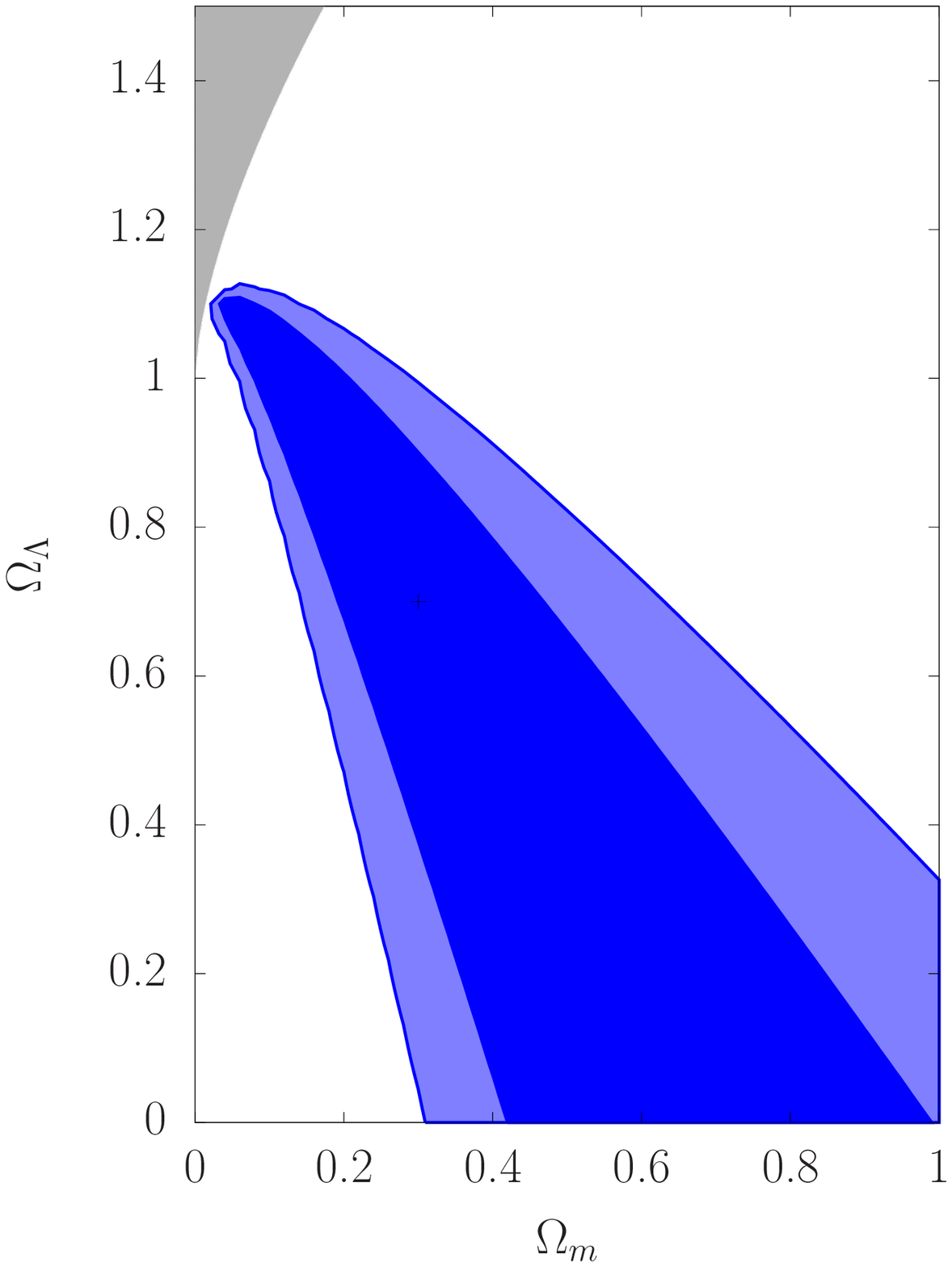}
  \includegraphics[width = 0.19 \textwidth]{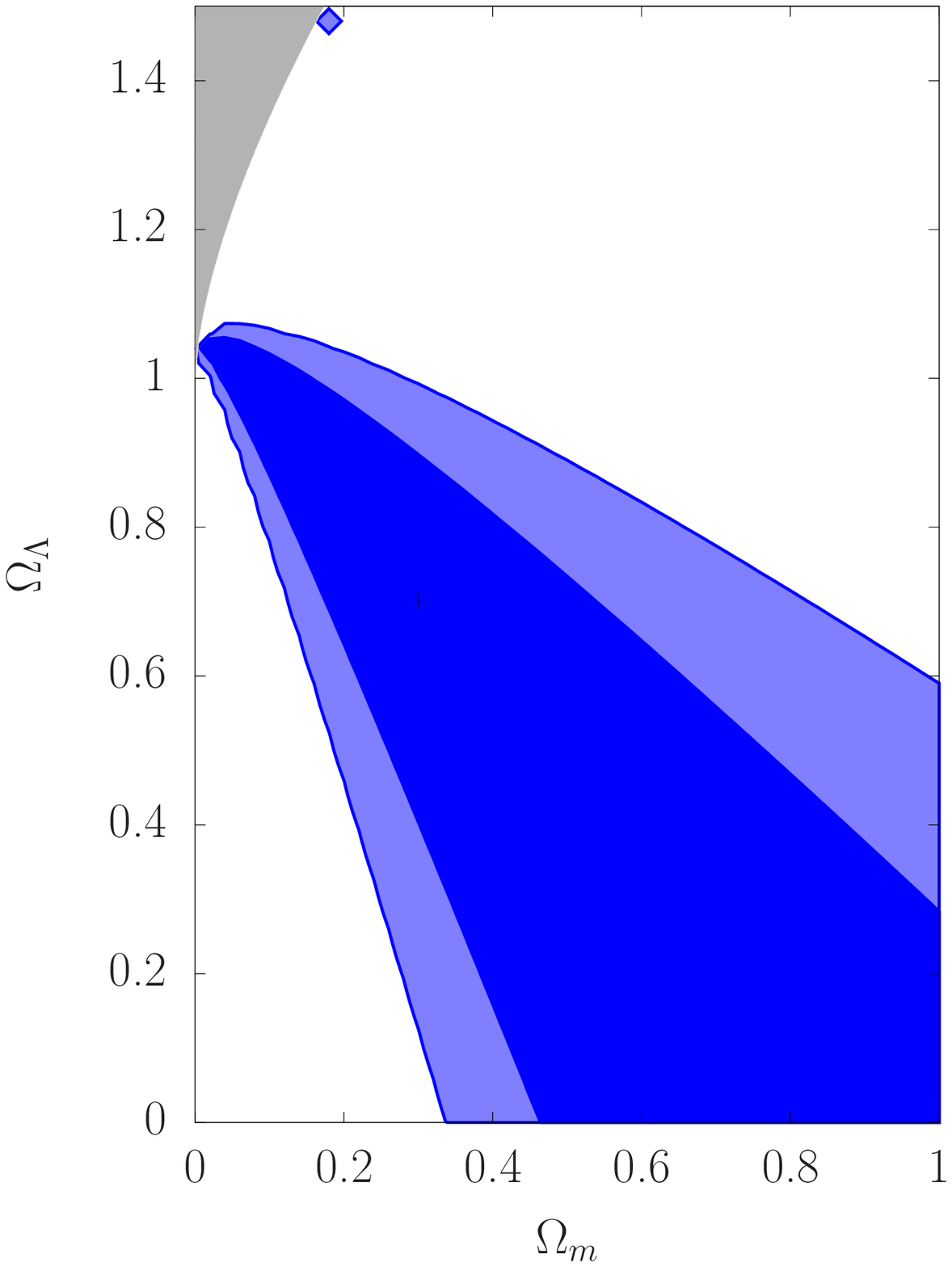}
  \includegraphics[width = 0.19 \textwidth]{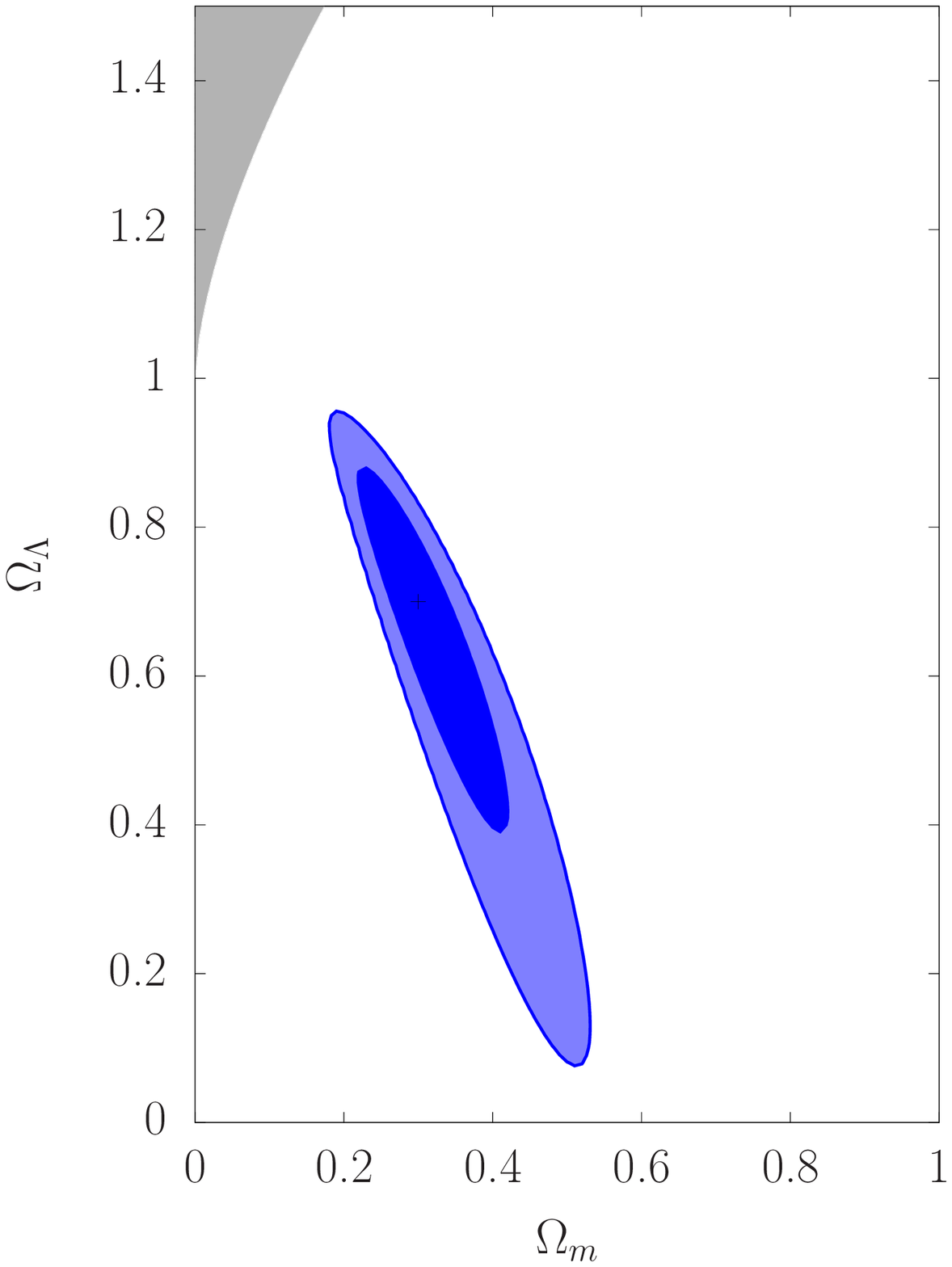}
  \\
  \includegraphics[width = 0.19 \textwidth]{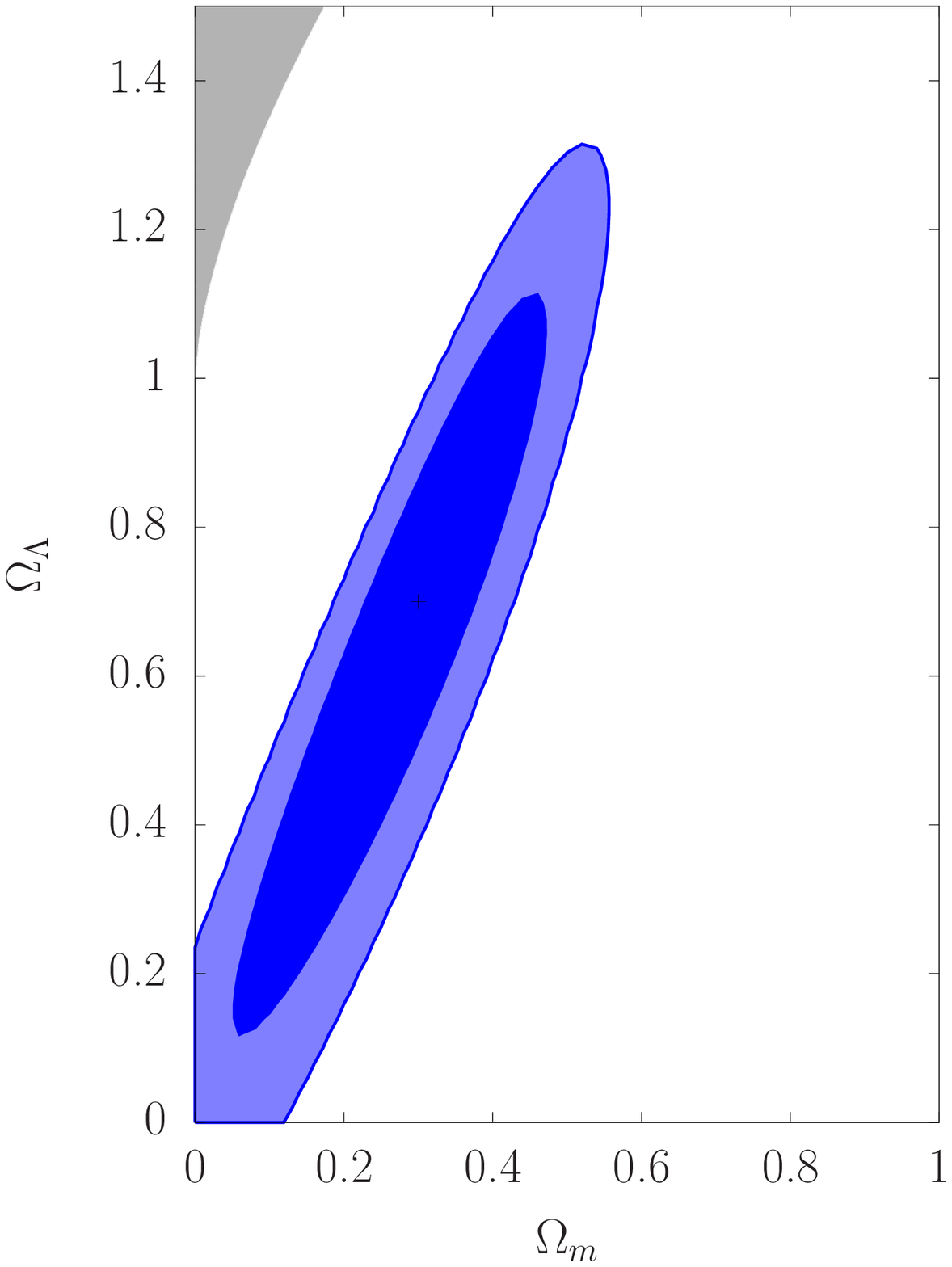}
  \includegraphics[width = 0.19 \textwidth]{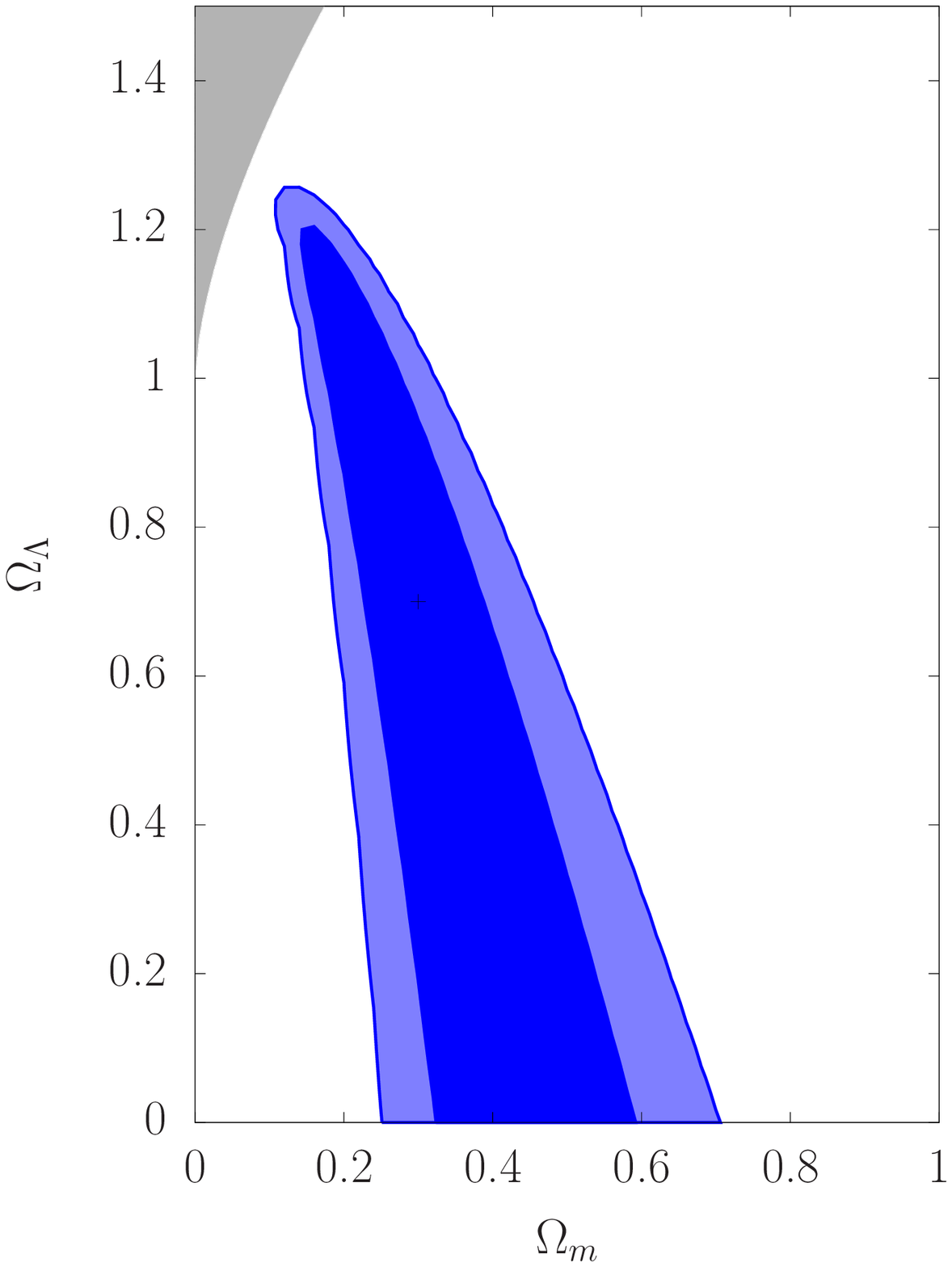}
  \includegraphics[width = 0.19 \textwidth]{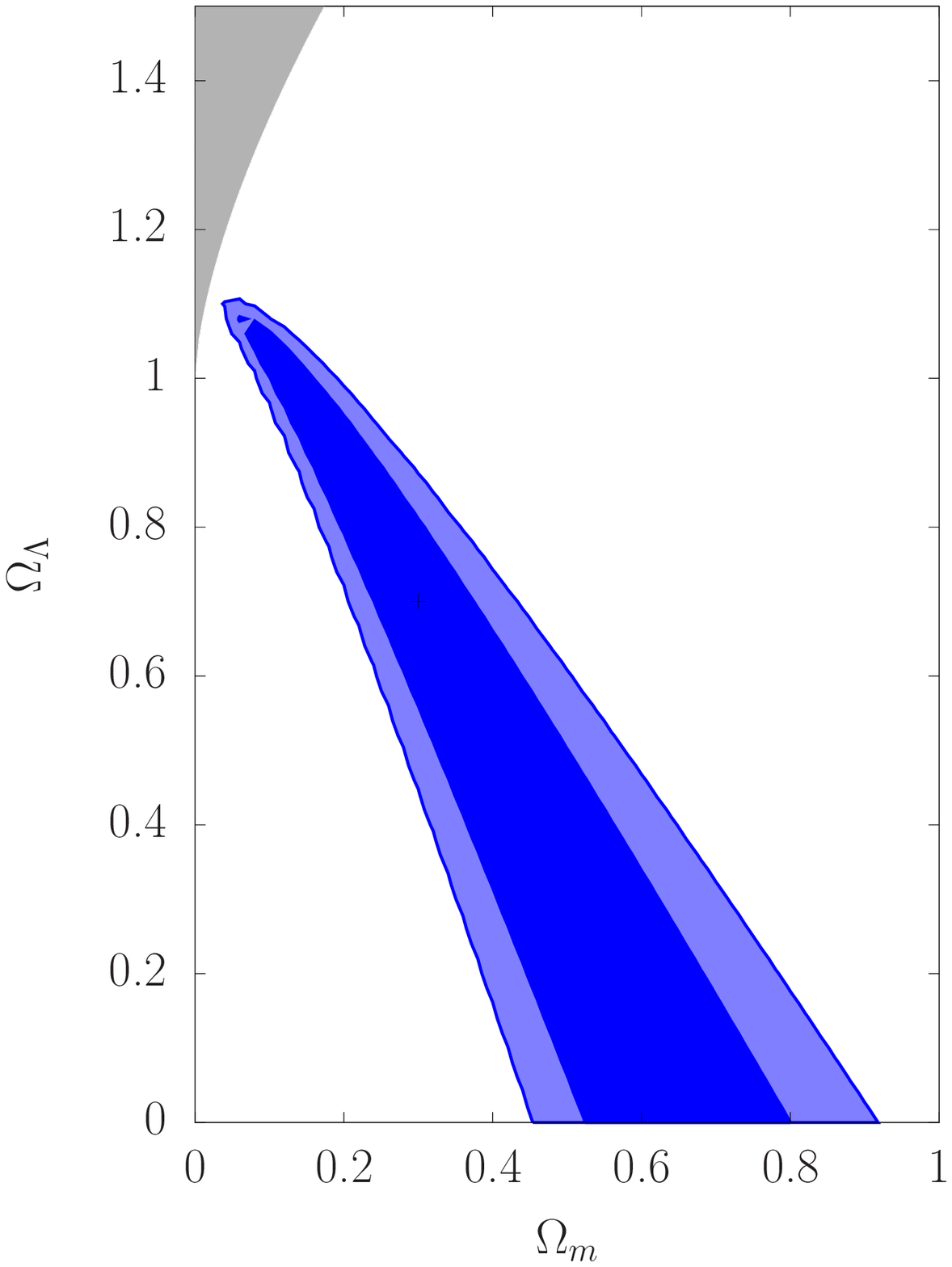}
  \includegraphics[width = 0.19 \textwidth]{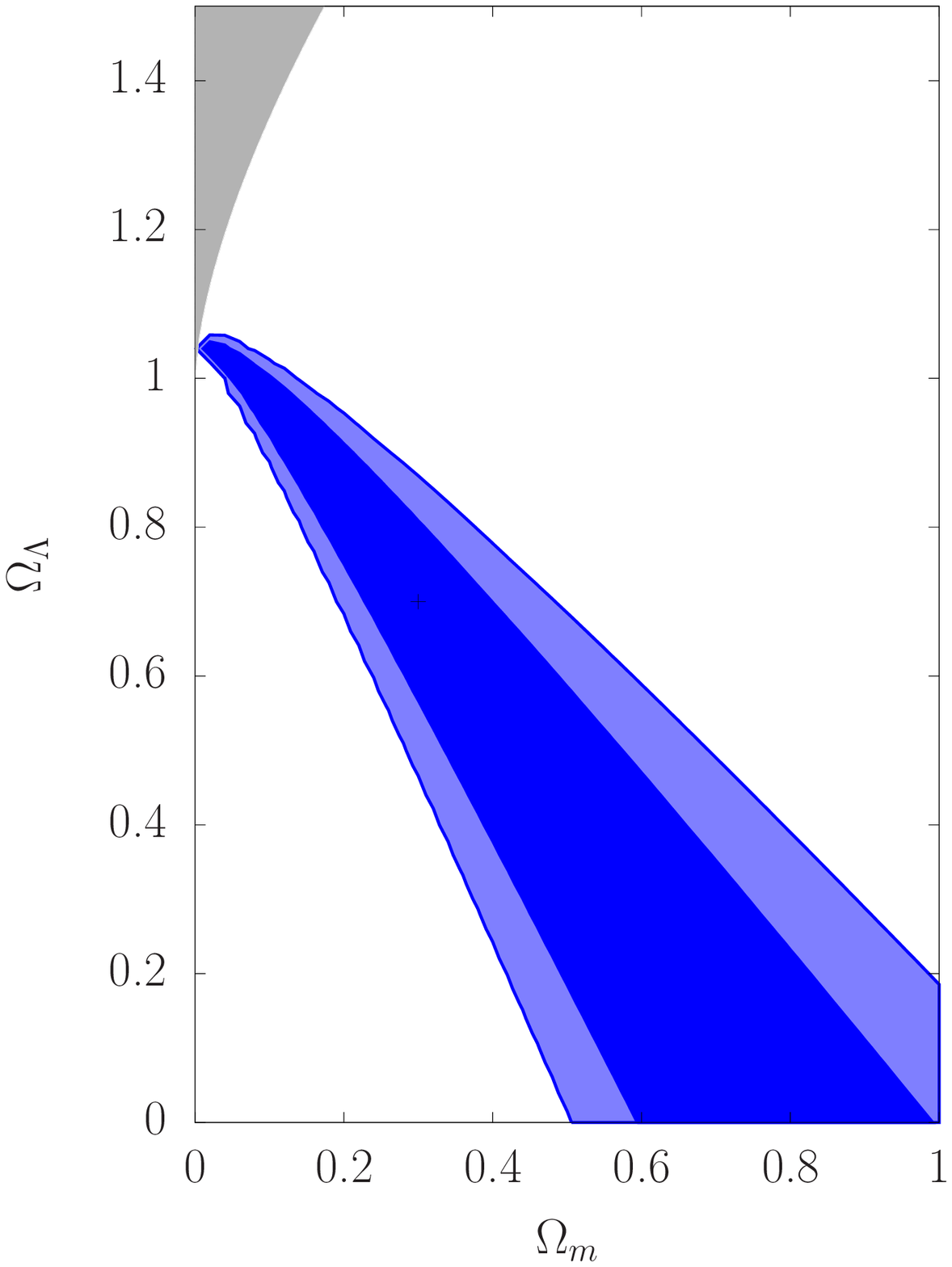}
  \includegraphics[width = 0.19 \textwidth]{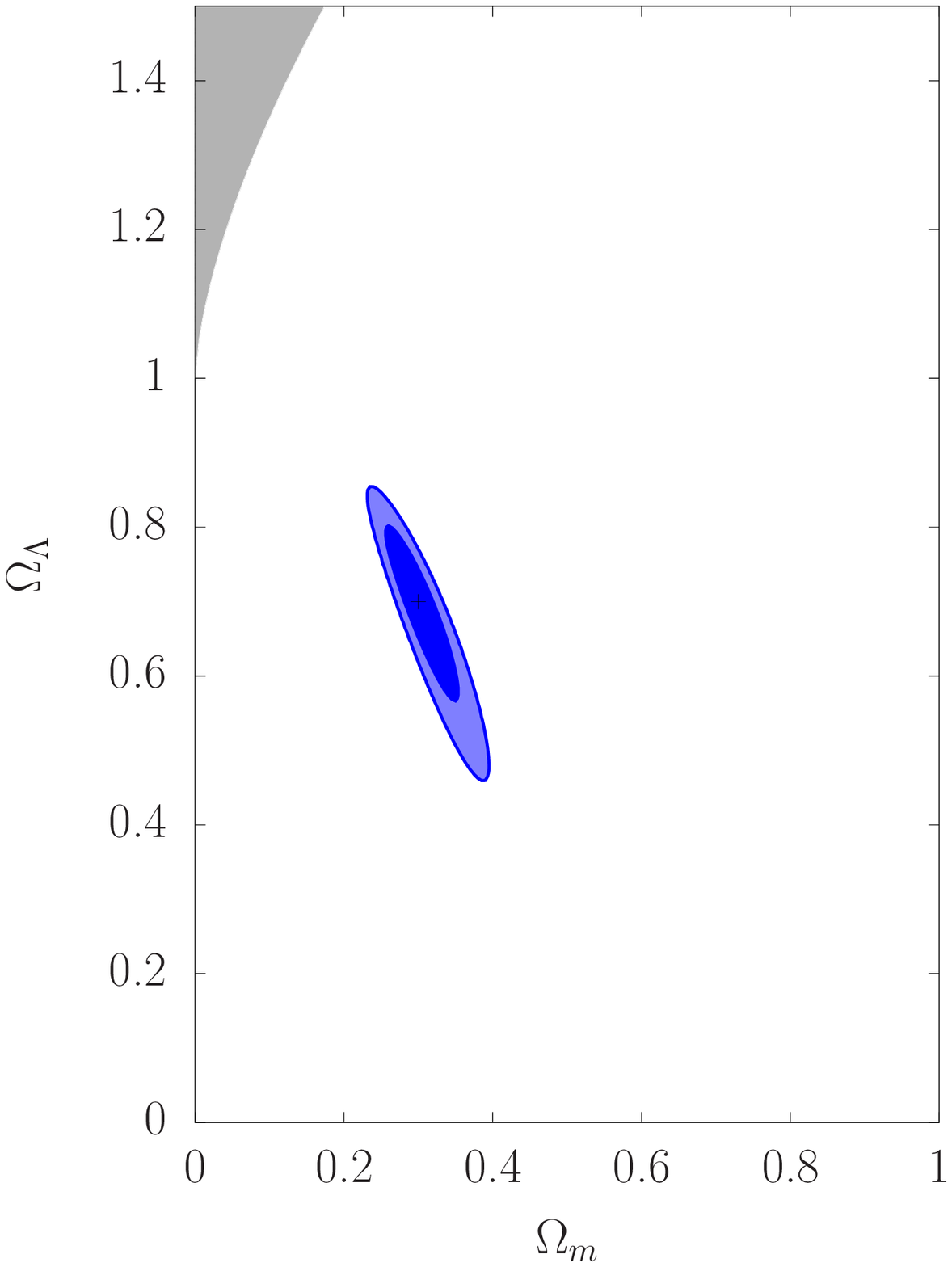}
  \caption
  {
    $68.3\%$ and $95.4\%$ confidence regions
    in the $(\Omega_m, \Omega_\Lambda)$ plane
    for the $\Lambda$CDM model.
    The flat $\Lambda$CDM with $\Omega_m = 0.3$ was used as the
    fiducial model and is represented by the black plus sign in the
    figures.
    The rows represent different intrinsic scatters of the luminosity
    relation for the standard candles.
    From top to bottom,
    $\sigma_{\mathrm{int}, 0} = 0.4$, $0.3$, $0.2$, $0.1$.
    The columns represent different redshift distributions of standard
    candles.
    From left to right, $500$ standard candles uniformly distributing
    in the redshift range $[0.1, 1]$, $[1, 2]$, $[2, 4]$,
    $[4, 7]$, $[0.1, 7]$ were used.
    The luminosity relation was assumed to have only one luminosity
    indicator involved.
    The upper left gray region in the figures represent the parameter
    space for which the universe does not experience a big bang in the
    past.
  }
  \label{fig:mx}
\end{figure*}

\begin{figure*}[tbp]
  \centering
  \includegraphics[width = 0.19 \textwidth]{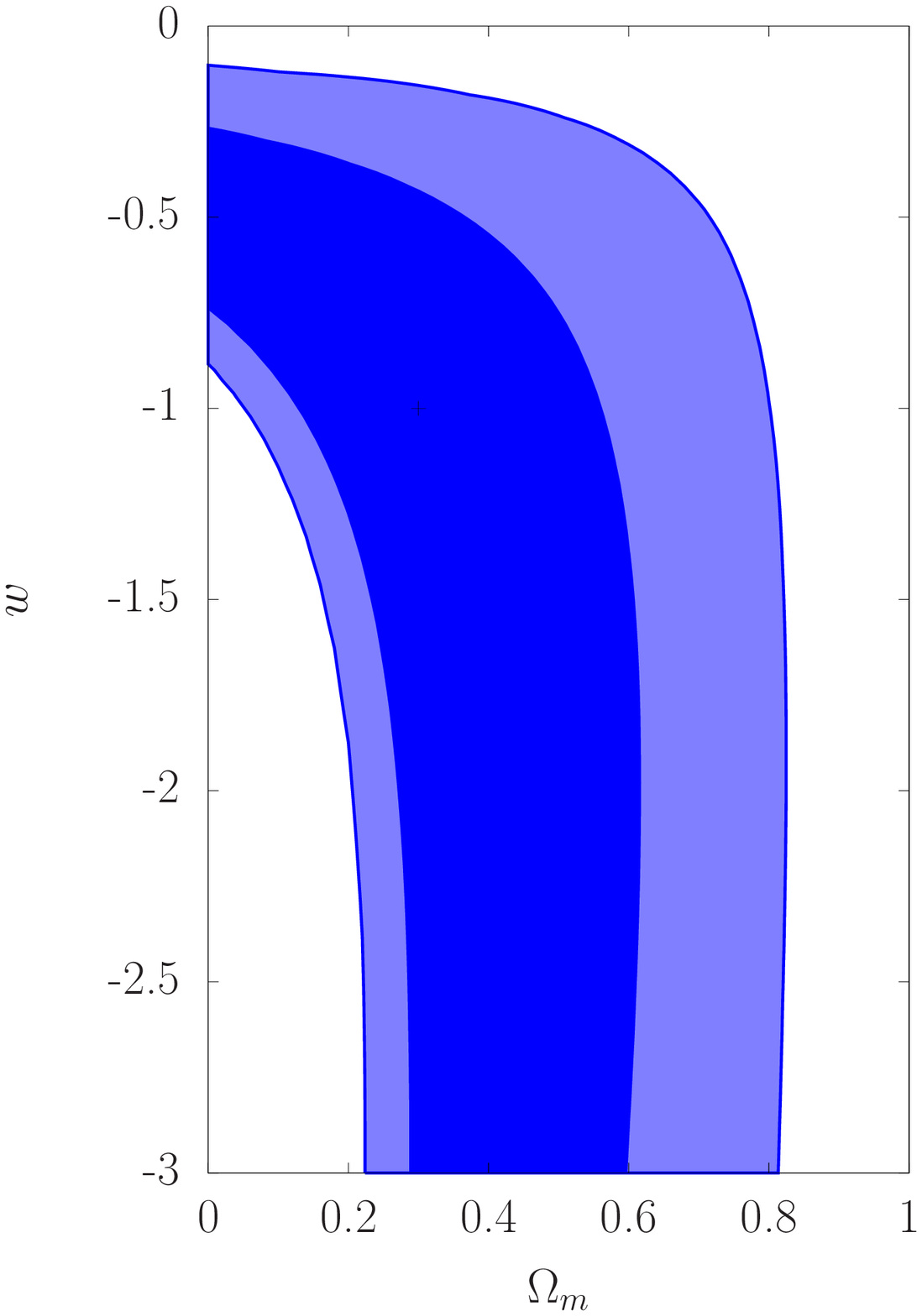}
  \includegraphics[width = 0.19 \textwidth]{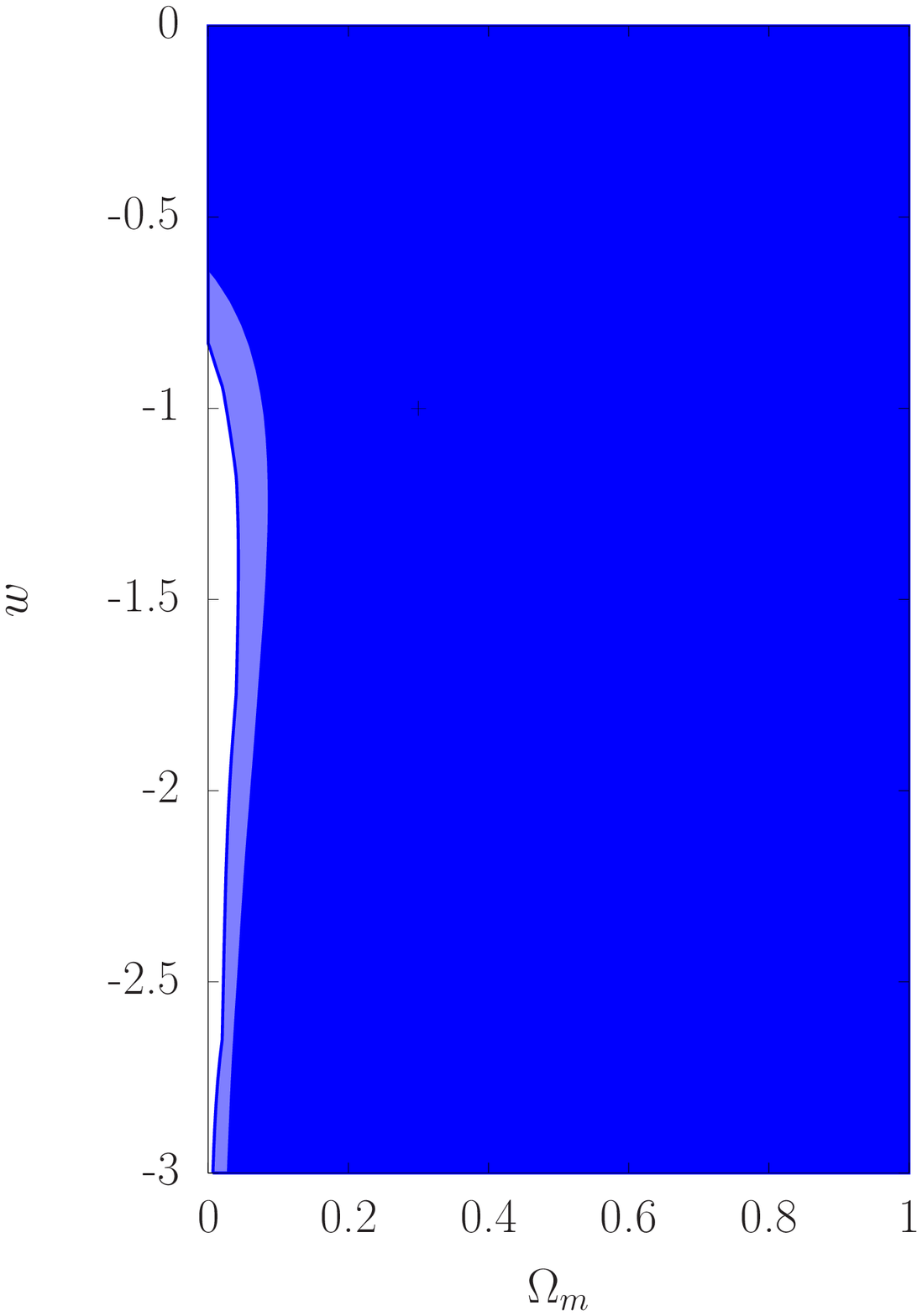}
  \includegraphics[width = 0.19 \textwidth]{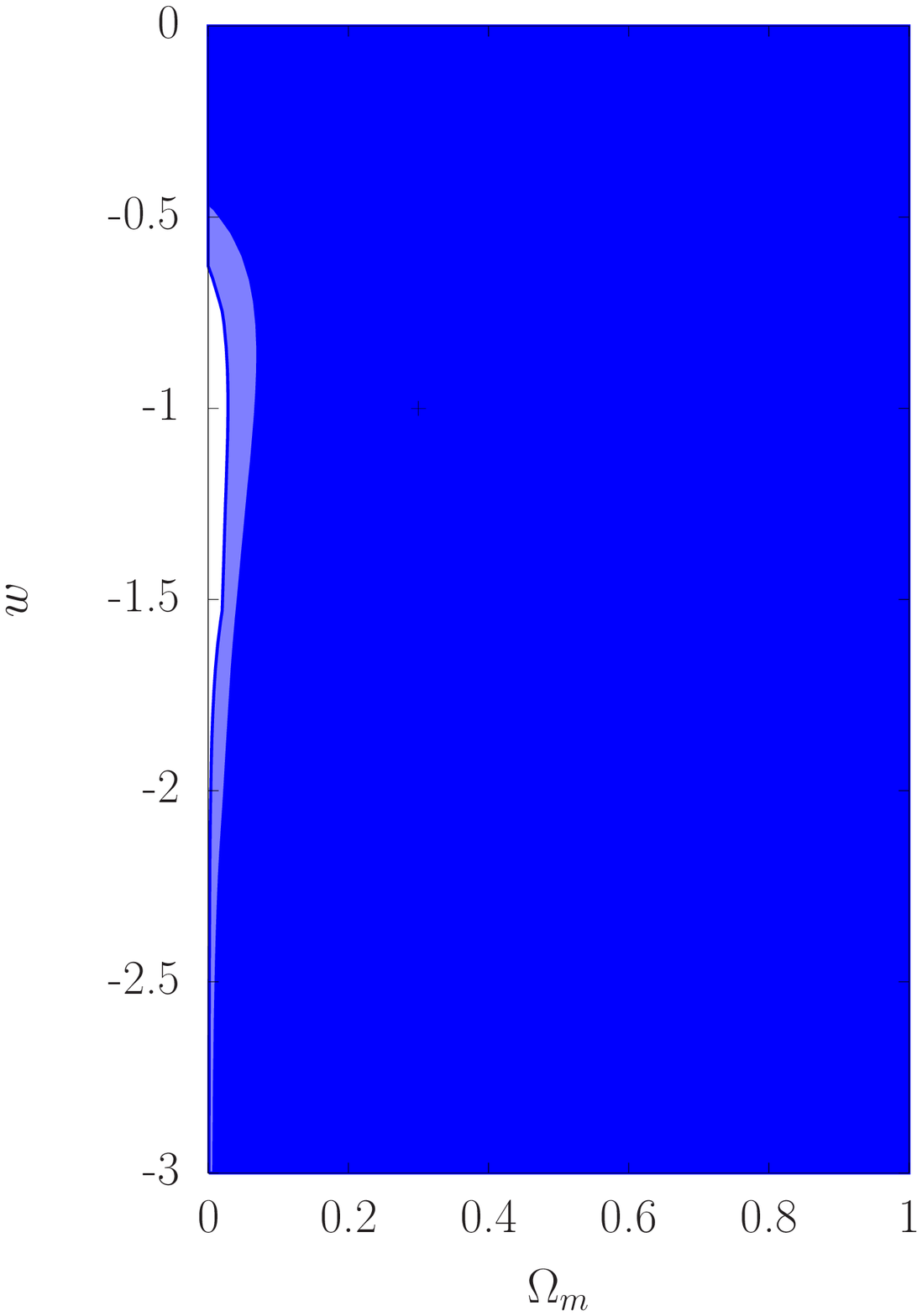}
  \includegraphics[width = 0.19 \textwidth]{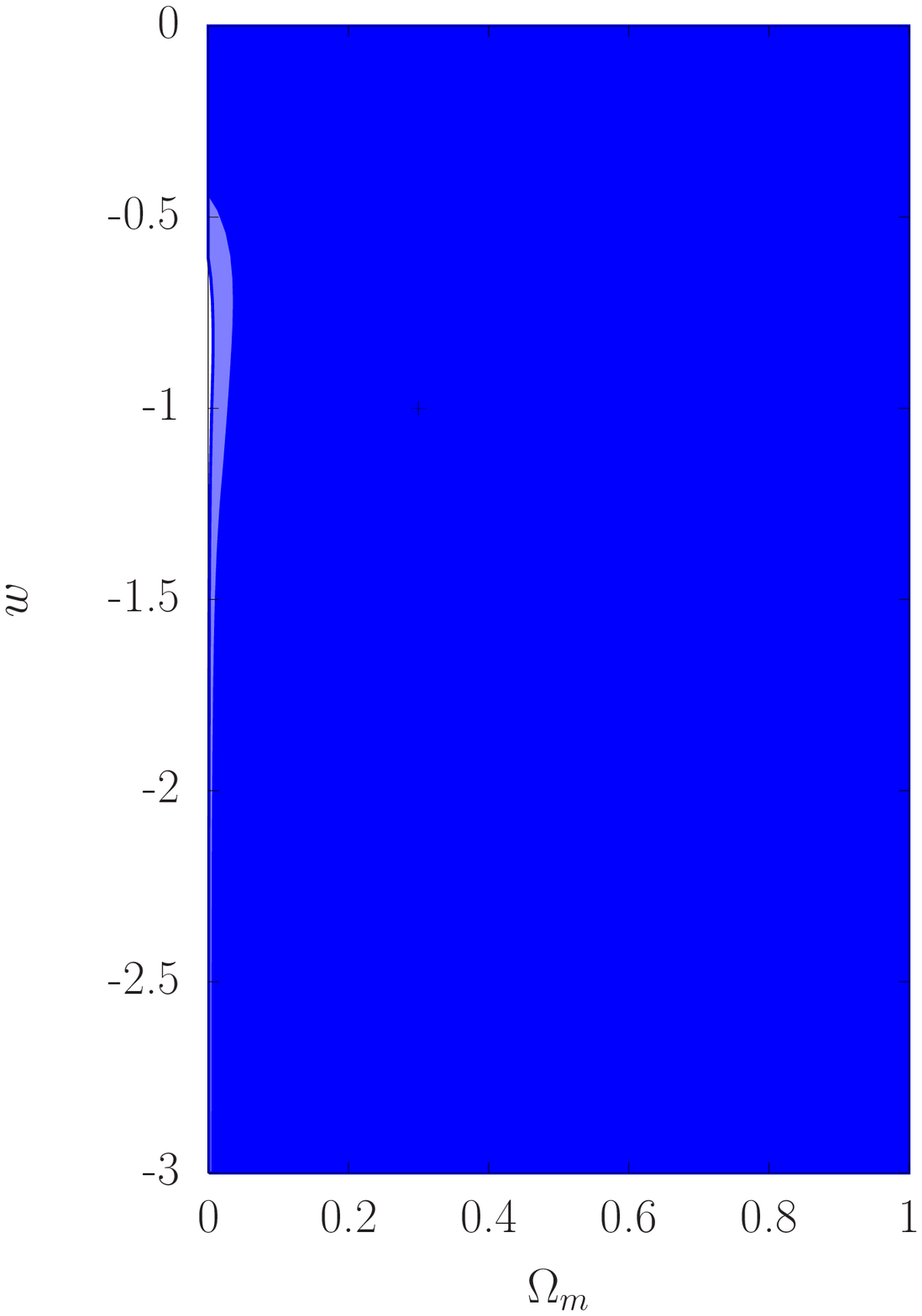}
  \includegraphics[width = 0.19 \textwidth]{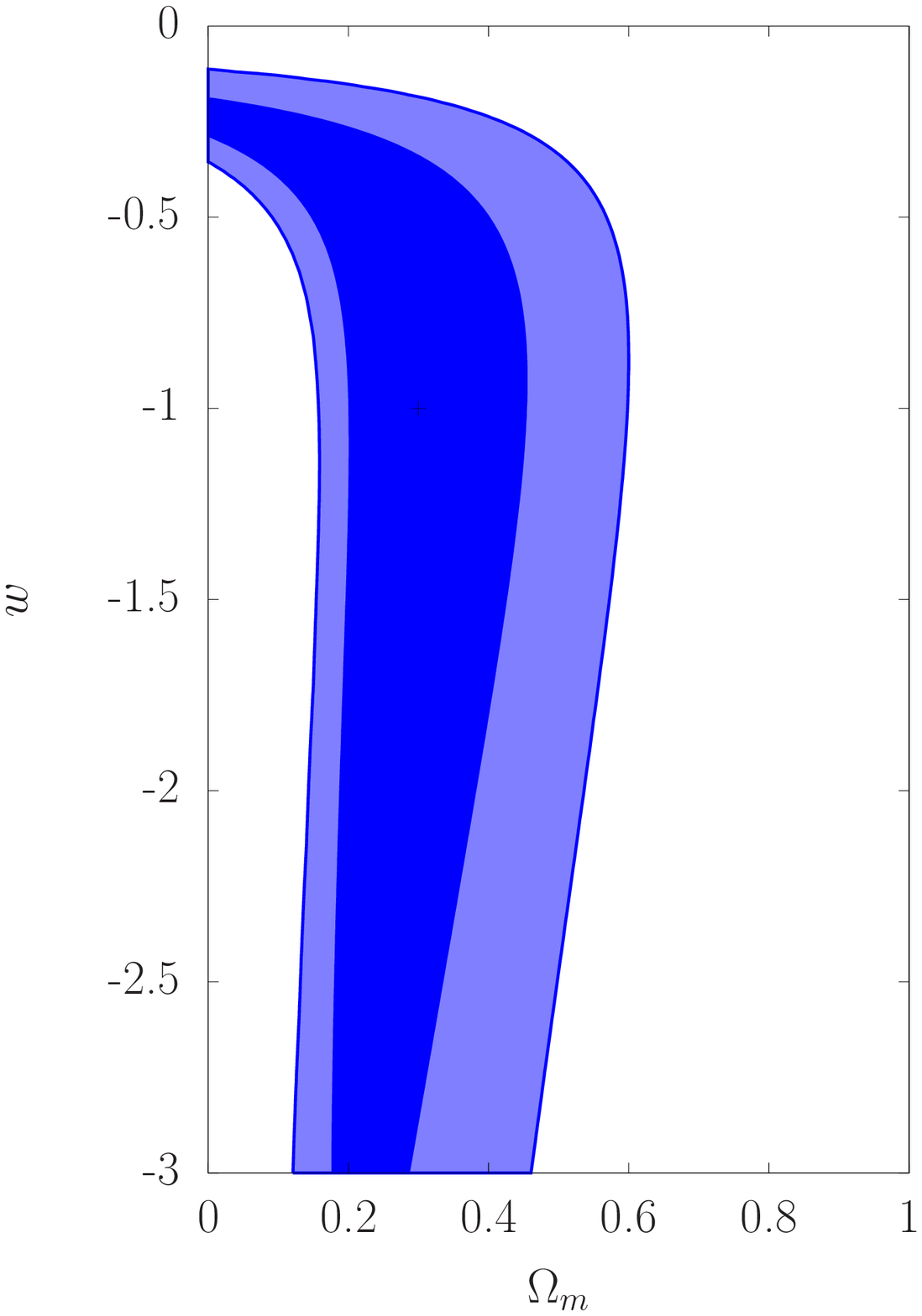}
  \\
  \includegraphics[width = 0.19 \textwidth]{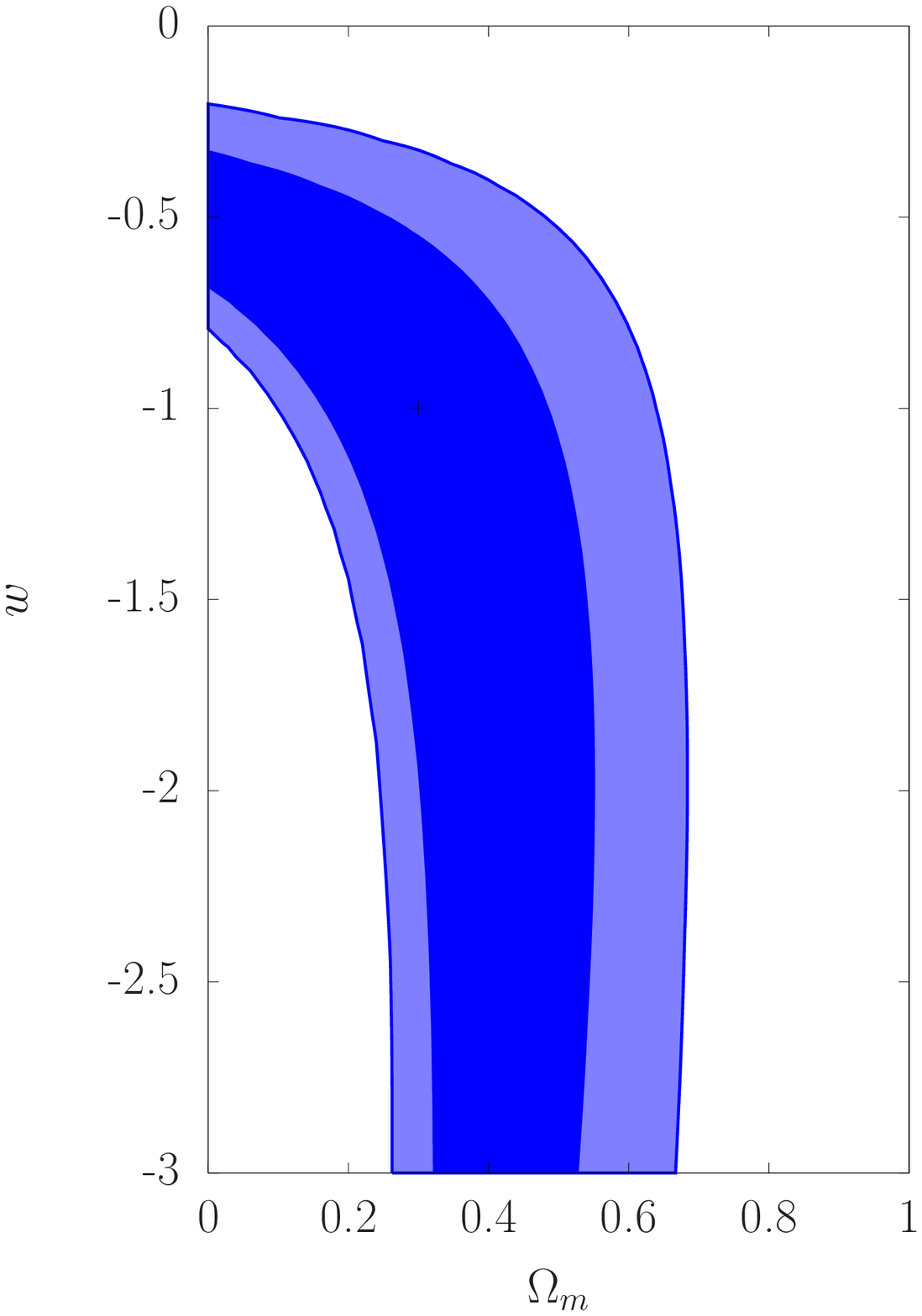}
  \includegraphics[width = 0.19 \textwidth]{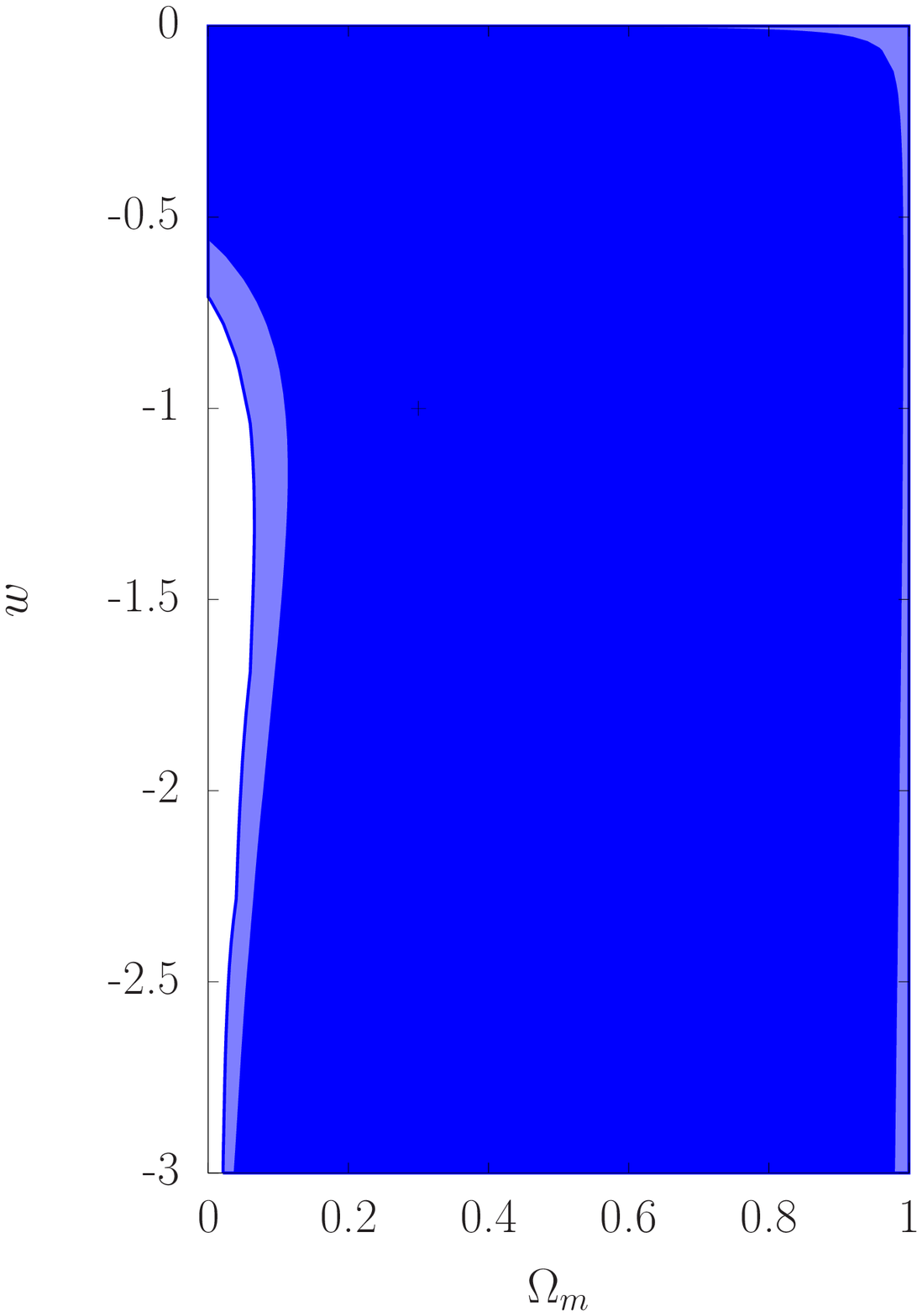}
  \includegraphics[width = 0.19 \textwidth]{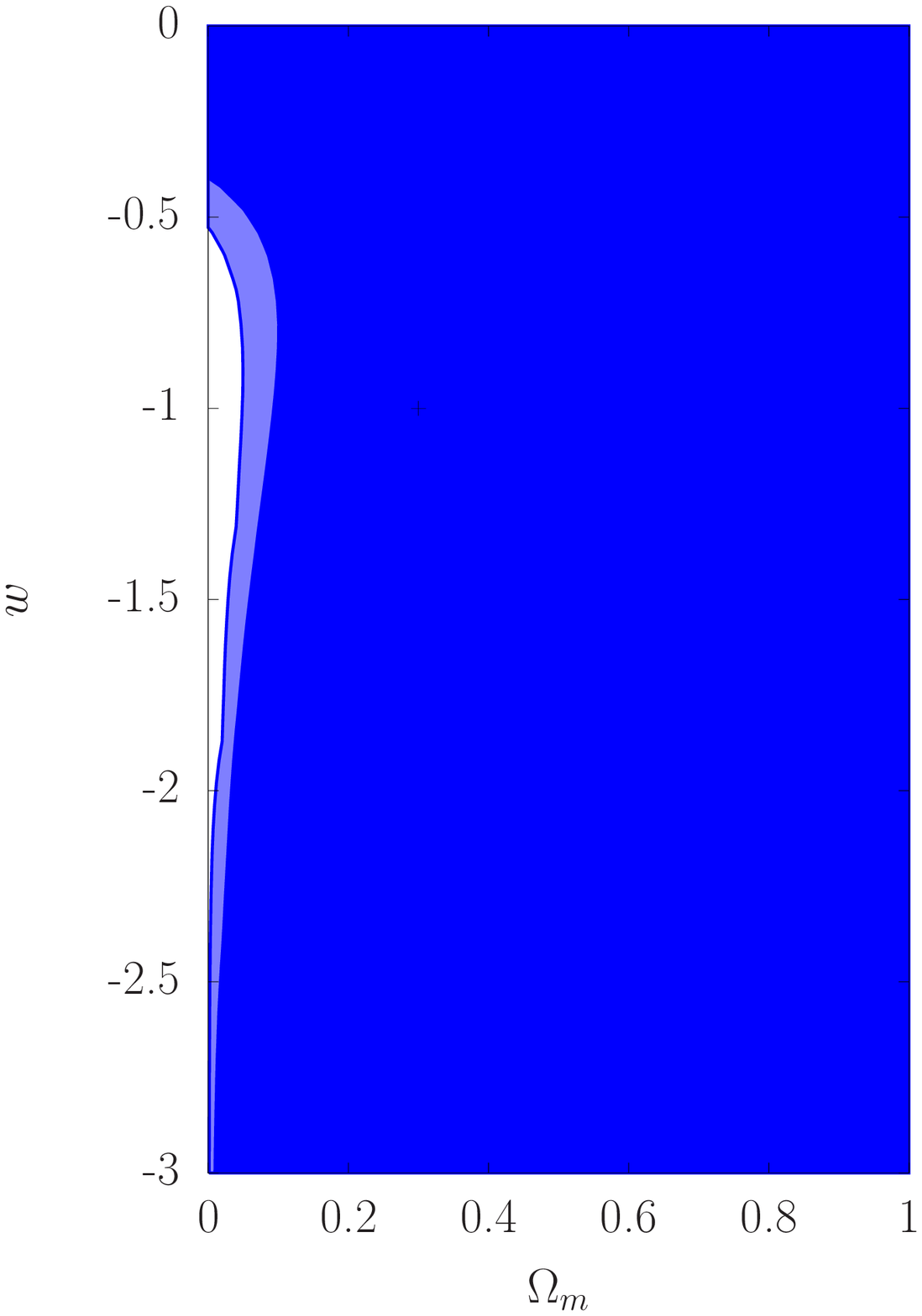}
  \includegraphics[width = 0.19 \textwidth]{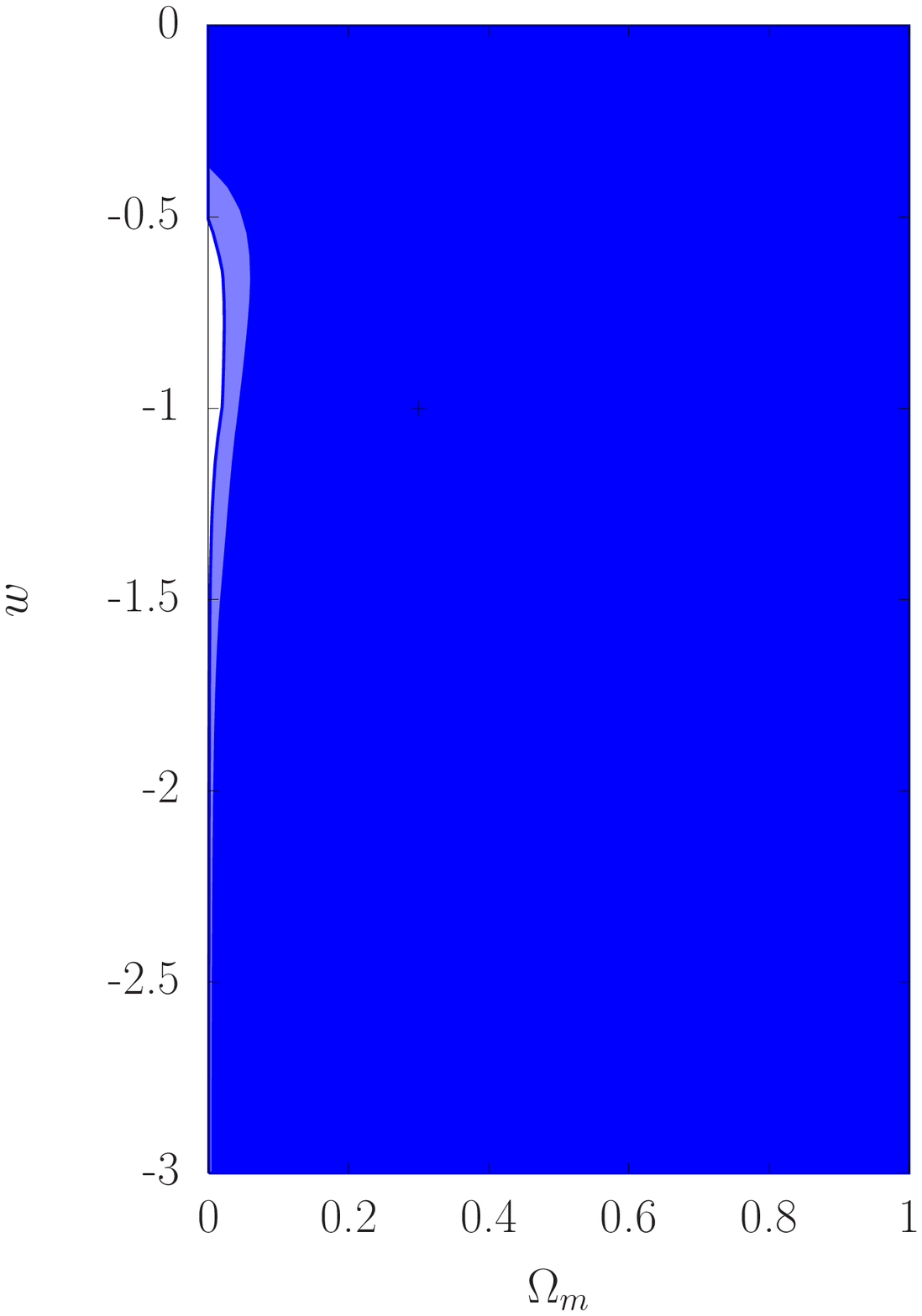}
  \includegraphics[width = 0.19 \textwidth]{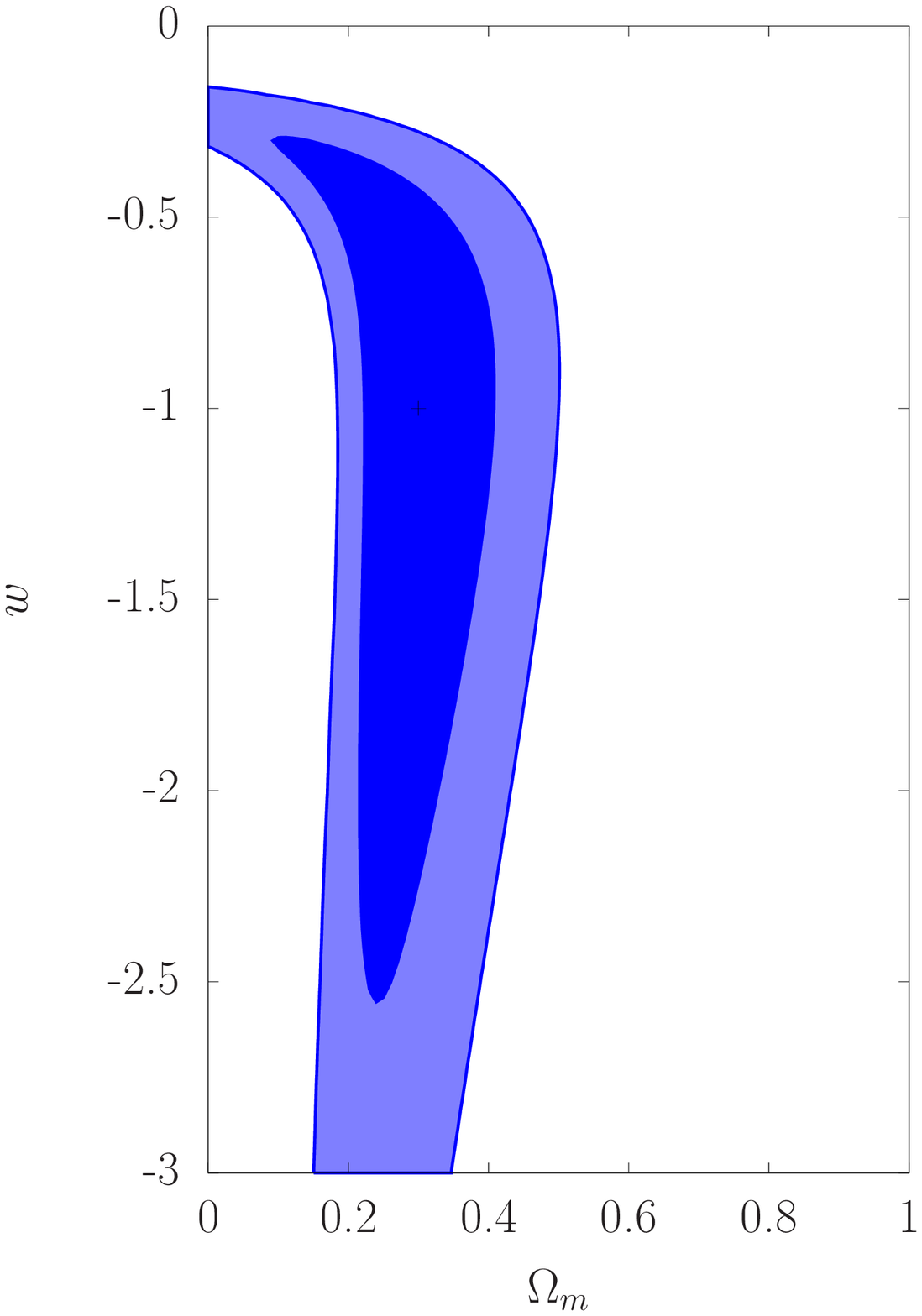}
  \\
  \includegraphics[width = 0.19 \textwidth]{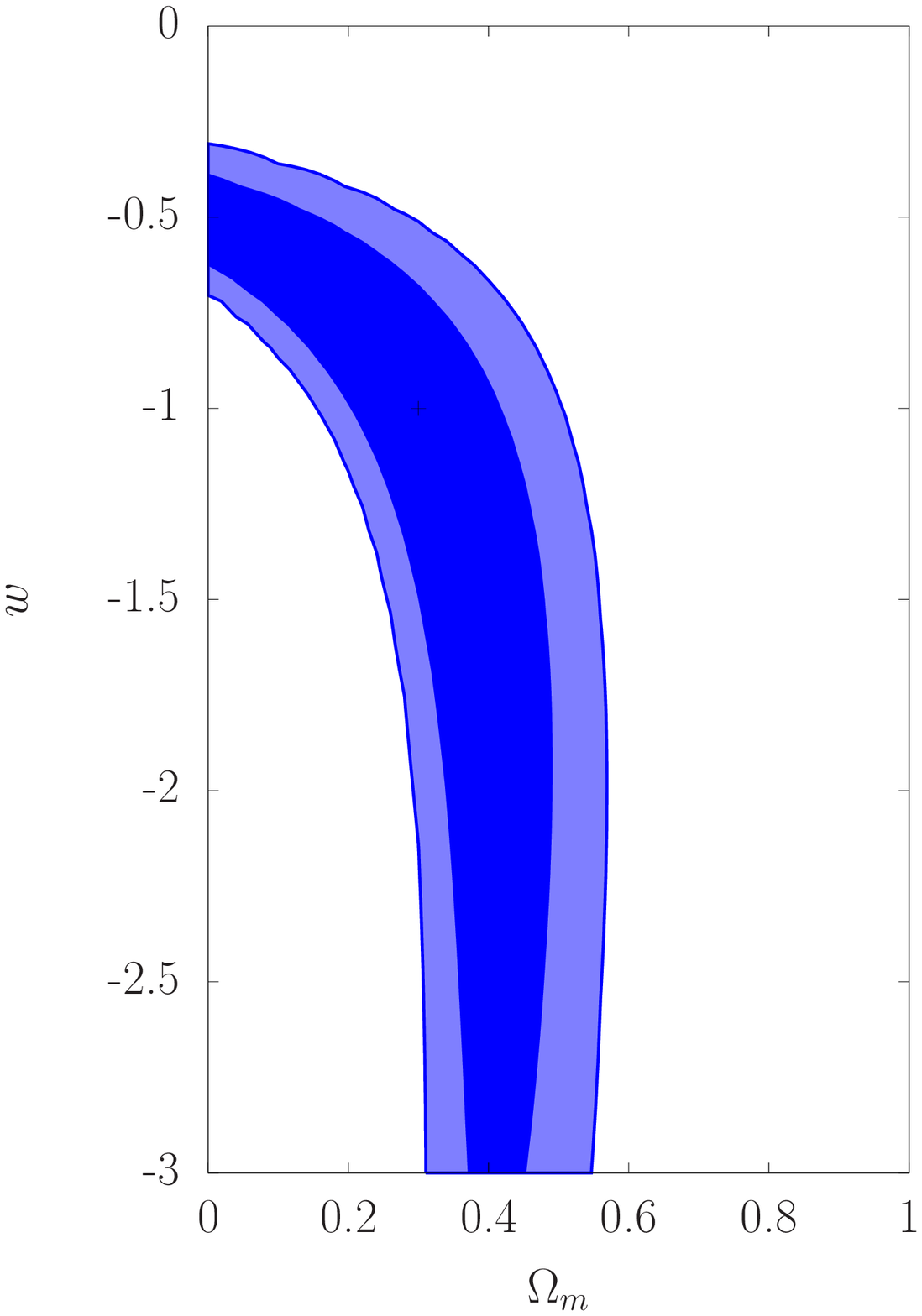}
  \includegraphics[width = 0.19 \textwidth]{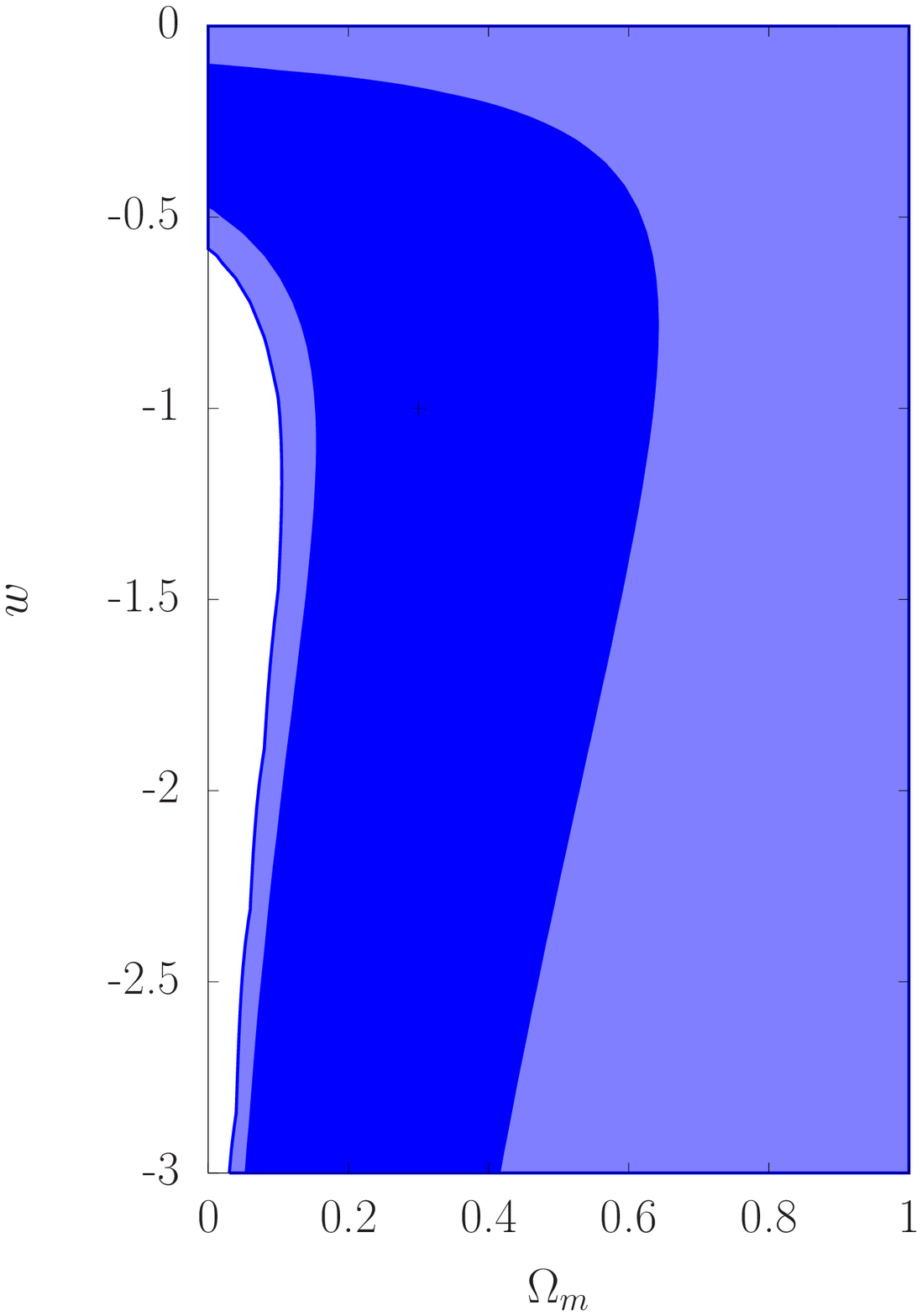}
  \includegraphics[width = 0.19 \textwidth]{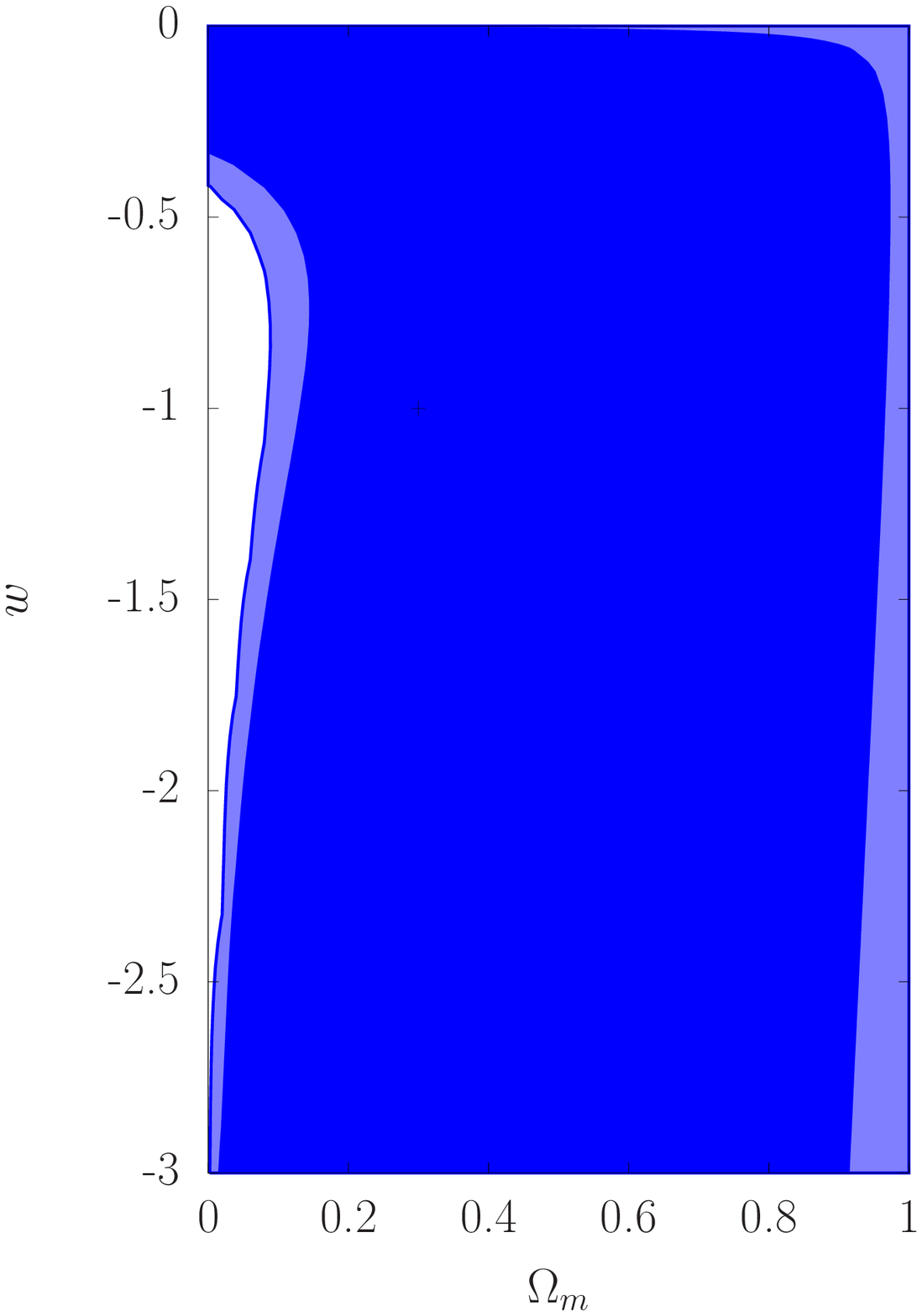}
  \includegraphics[width = 0.19 \textwidth]{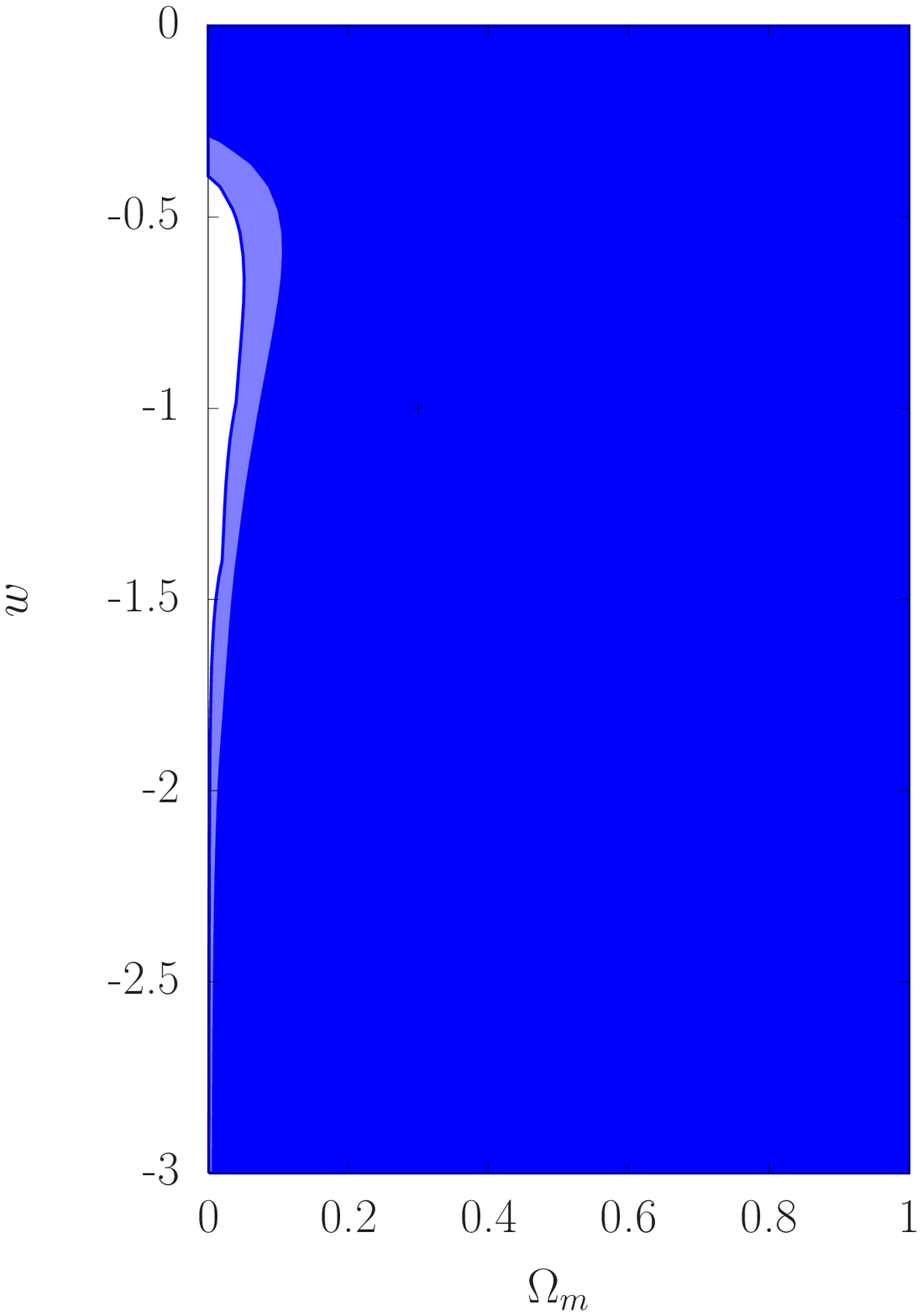}
  \includegraphics[width = 0.19 \textwidth]{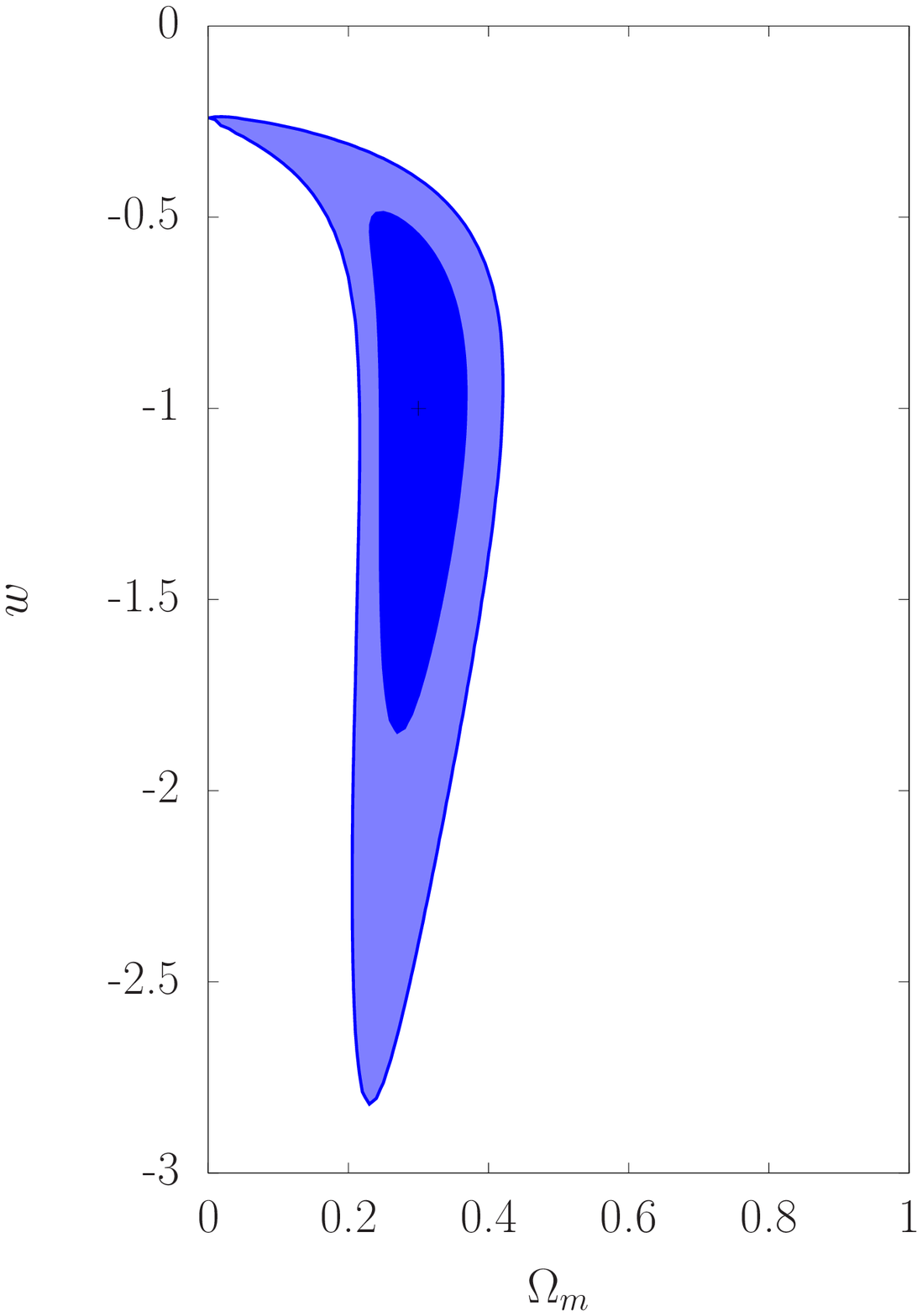}
  \\
  \includegraphics[width = 0.19 \textwidth]{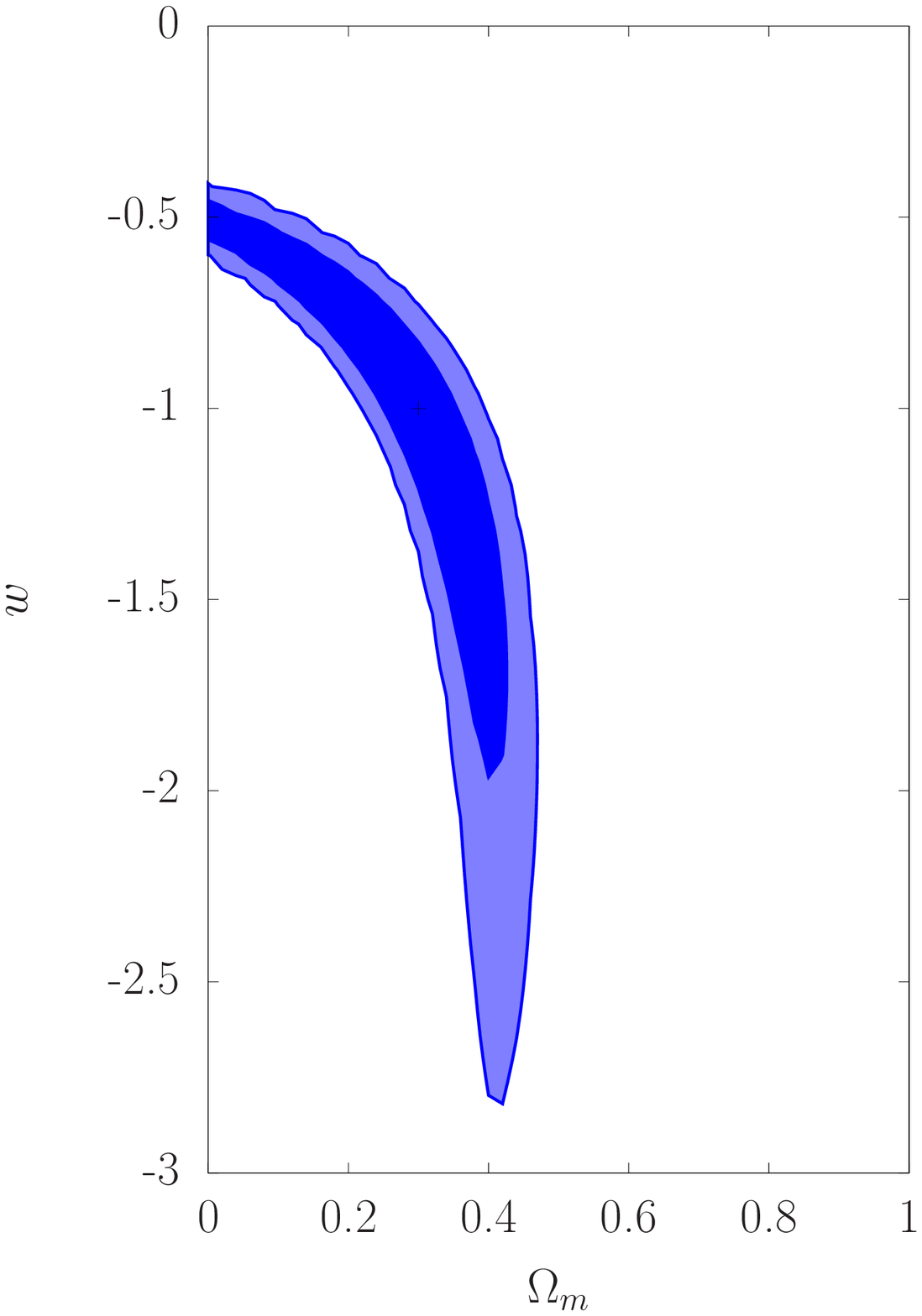}
  \includegraphics[width = 0.19 \textwidth]{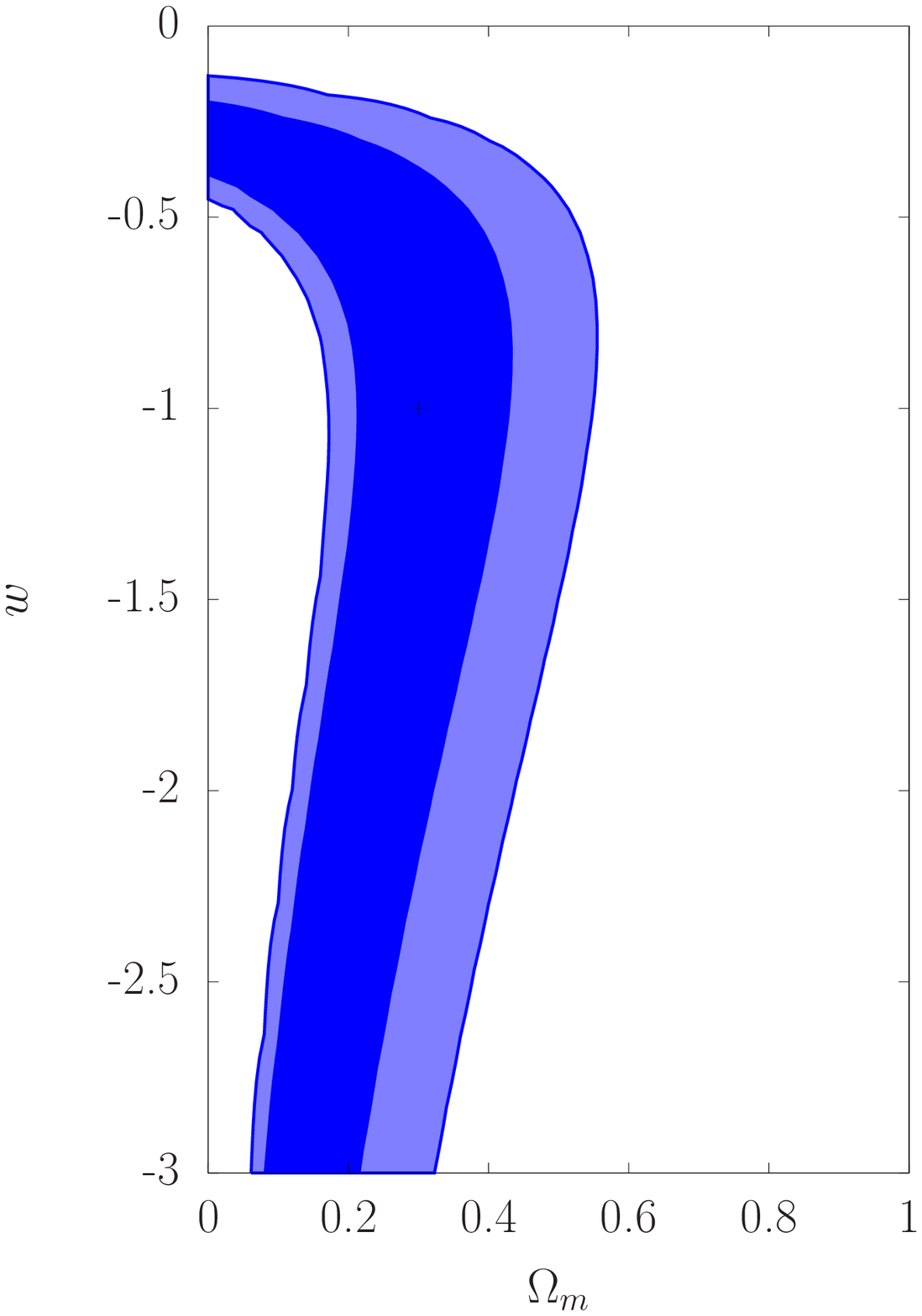}
  \includegraphics[width = 0.19 \textwidth]{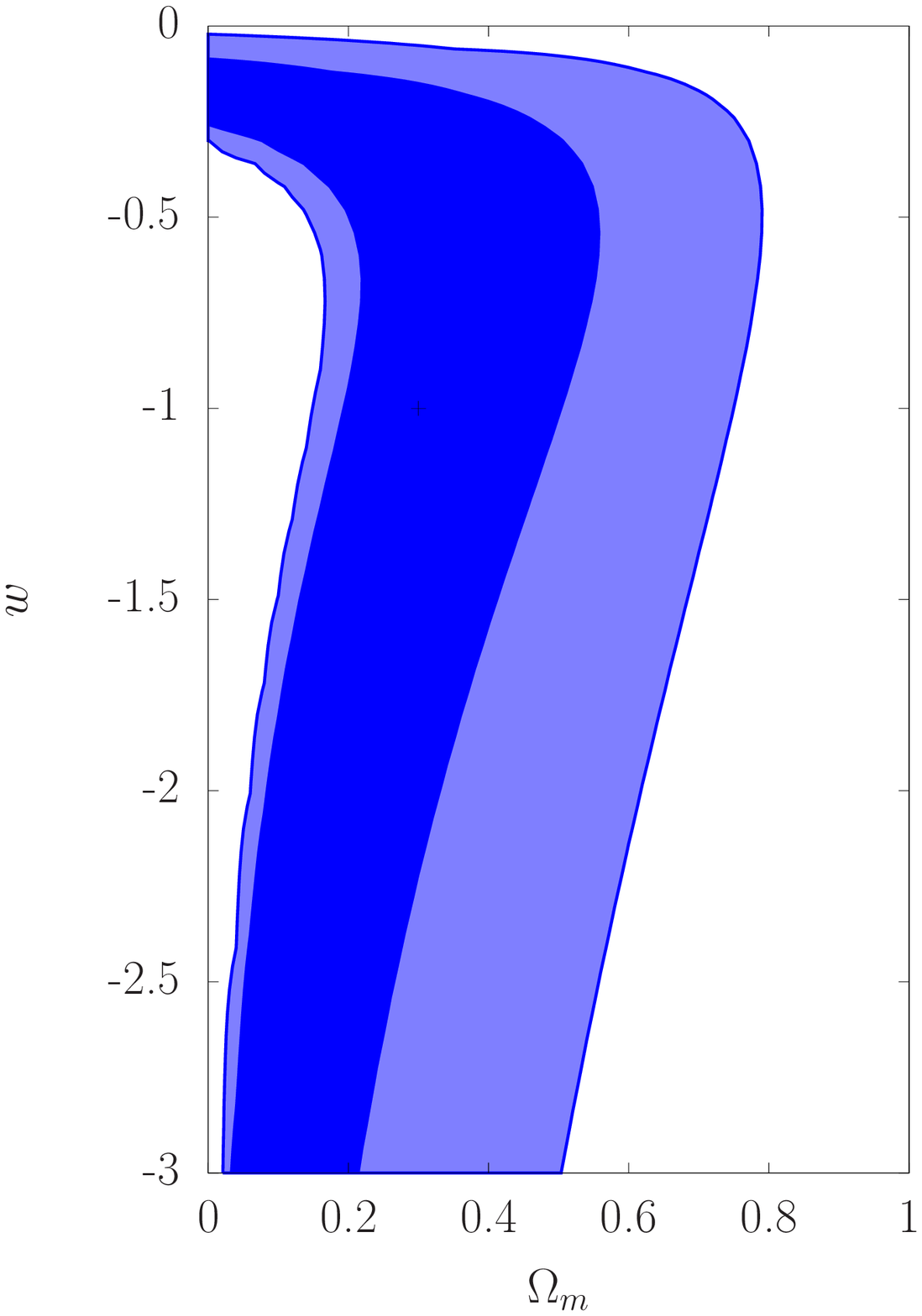}
  \includegraphics[width = 0.19 \textwidth]{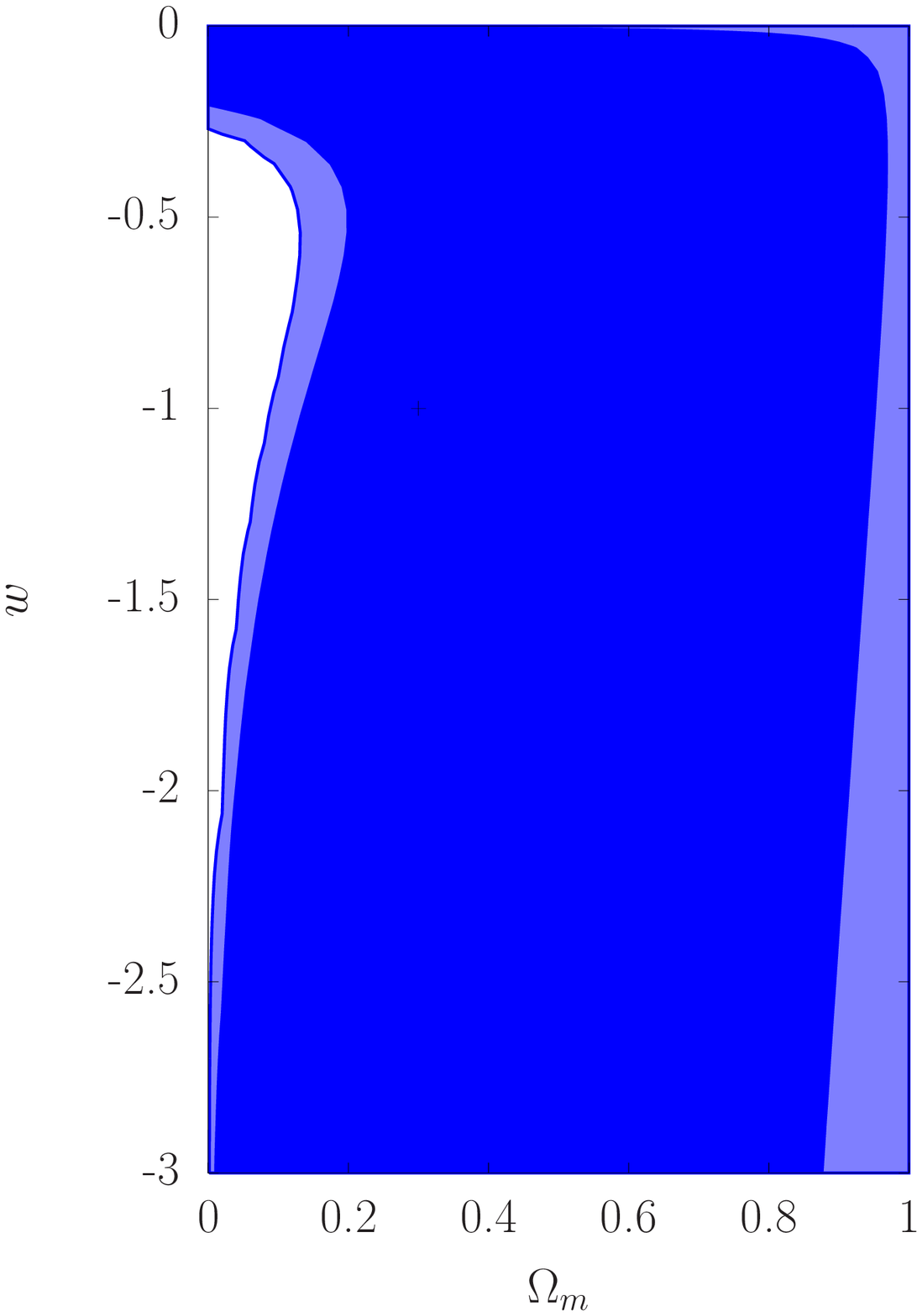}
  \includegraphics[width = 0.19 \textwidth]{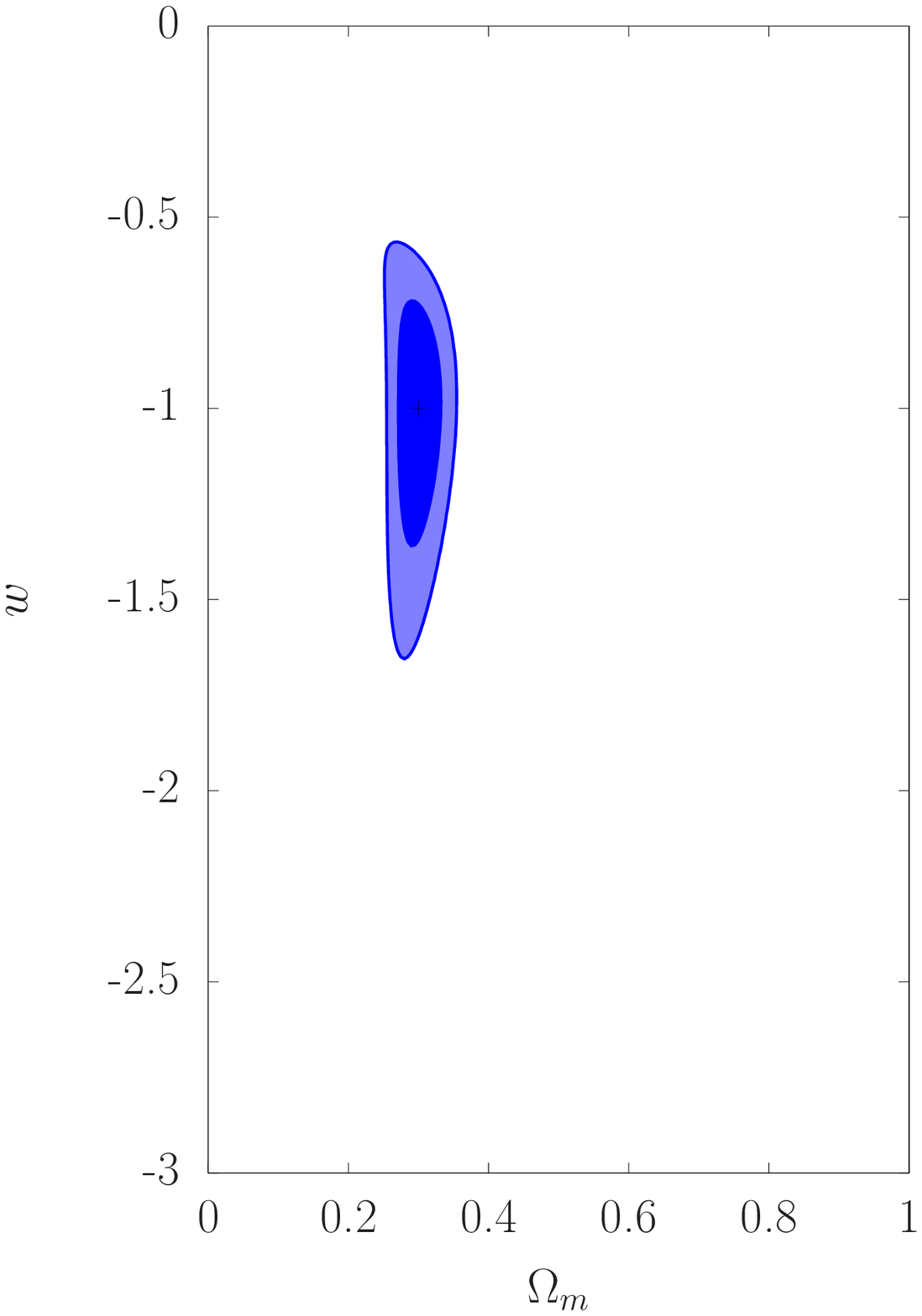}
  \caption
  {
    $68.3\%$ and $95.4\%$ confidence regions
    in the $(\Omega_m, w)$ plane
    for the flat $w$CDM model.
    The flat $\Lambda$CDM with $\Omega_m = 0.3$ was used as the
    fiducial model and is represented by the black plus sign in the
    figures.
    The rows represent different intrinsic scatters of the luminosity
    relation for the standard candles.
    From top to bottom,
    $\sigma_{\mathrm{int}, 0} = 0.4$, $0.3$, $0.2$, $0.1$.
    The columns represent different redshift distributions of standard
    candles.
    From left to right, $500$ standard candles uniformly distributing
    in the redshift range $[0.1, 1]$, $[1, 2]$, $[2, 4]$,
    $[4, 7]$, $[0.1, 7]$ were used.
    The luminosity relation was assumed to have only one luminosity
    indicator involved.
  }
  \label{fig:mw}
\end{figure*}

Here, utilizing Eqs.~(\ref{eq:lztheta_dl}) and (\ref{eq:Ltheta}), the
constraining capability of standard candles on cosmological parameters
was estimated using mock data.
The flat $\Lambda$CDM with $\Omega_m = 0.3$ was used as the fiducial
model and $p = 3$ was assumed.
The constraints on $(\Omega_m, \Omega_\Lambda)$ for the $\Lambda$CDM
model and on $(\Omega_m, w)$ for the flat $w$CDM model were studied.
The dependence of the constraints on the intrinsic scatter of the
luminosity relation and the redshift distribution of the standard
candles was specifically investigated.
For the intrinsic scatter of the luminosity relation,
$\sigma_{\mathrm{int}, 0} = 0.4$, $0.3$, $0.2$, and $0.1$
were considered.
For the redshift distribution of the standard candles,
five cases were considered, i.e., $500$ standard candles uniformly
distributing in the redshift range
$[0.1, 1]$, $[1, 2]$, $[2, 4]$, $[4, 7]$, and $[0.1, 7]$.
These values were chosen specifically by keeping in mind the current
development of the GRBs as standard candles
(see e.g.~\cite{Qi:2011tr}).
The results are presented in Figs.~\ref{fig:mx} and~\ref{fig:mw}.

From Figs.~\ref{fig:mx} and~\ref{fig:mw}, we can see that,
as expected,
the smaller the intrinsic scatter $\sigma_{\mathrm{int}, 0}$ is,
the tighter the constraints are.
For the dependence of the constraints on the redshift distribution of
the standard candles, we can see, especially from Fig.~\ref{fig:mx},
that the constraints from standard candles of different redshifts show
different degeneracy orientations, thus a combination of standard
candles from a wide redshift range can self break the degeneracy and
improve the constraints significantly.
This is because that $\mathcal{L}(\theta)$ depends on
$l(z, \theta, \theta_0)$ through its variance along the redshift.
Standard candles with a narrow redshift distribution can only reflect
the local redshift variance of $l(z, \theta, \theta_0)$,
and the redshift variance of $l(z, \theta, \theta_0)$ has different
features at different redshifts,
which leads to different degeneracy orientations of
$\mathcal{L}(\theta)$.
In contrast, standard candles with a wide redshift distribution can
reflect the global redshift variance of $l(z, \theta, \theta_0)$,
the derived $\mathcal{L}(\theta)$ is more sensitive to the variation
of the cosmological parameters $\theta$,
which means tighter constraints.
Such a feature of standard candles was also shown
in~\cite{Ghisellini:2005vk, Firmani:2006hq}
by using stripes of constant $d_L$ at different redshifts in the
cosmological parameter space.

The self degeneracy breaking feature of a wide redshift distribution
of standard candles means that redshift distribution of standard
candles can play a similar role as the intrinsic scatter of the
luminosity relation in determining the tightness of the constraints.
For a given intrinsic scatter, standard candles of same sample size
with wider redshift distributions give tighter constraints.
A loose luminosity relation combined with a wide redshift distribution
of standard candles can have comparable tightness of constraint with
a tight luminosity relation combined with a narrow redshift
distribution of standard candles.
For example, in Fig.~\ref{fig:mx}, the contour plots of the top right
($\sigma_{\mathrm{int}, 0} = 0.4$ and $500$ standard candles uniformly
distributing in the redshift range $[0.1, 7]$)
and the bottom left
($\sigma_{\mathrm{int}, 0} = 0.1$ and $500$ standard candles uniformly
distributing in the redshift range $[0.1, 1]$)
give comparable tightness of constraint.

The self degeneracy breaking due to the redshift distribution of
standard candles is less obvious in Fig.~\ref{fig:mw} than in
Fig.~\ref{fig:mx}.
This is because that the constraints on the dark energy EOS $w$ mainly
come from standard candles at redshifts less than, say, about $2$.
The universe is matter dominated at high redshifts, where dark energy
does not contribute much in determining the evolution of the
universe, so the constraints on the dark energy from standard candles
at high redshifts are very weak and are thus less helpful in the
degeneracy breaking.

As mentioned earlier, the intrinsic scatter of the luminosity relation
and the redshift distribution of the standard candles are chosen by
keeping in mind the current development of GRBs as standard candles.
From the results, we can conclude that, even with the current level of
tightness of known luminosity relations (see e.g.~\cite{Qi:2011tr}),
GRBs can give comparable tightness of constraint with SNe Ia on the
components of the universe as long as the redshifts of the GRBs are
diversifying enough.
However, for a substantial constraint on the dark energy EOS, we need
tighter luminosity relations for GRBs.

\begin{acknowledgments}
  This work was supported by the National Natural Science Foundation
  of China (Grant No.~10973039, 11203079, and 11373068),
  Key Laboratory of Dark Matter and Space Astronomy (Grant
  No.~DMS2011KT001), and the Project of Knowledge Innovation Program
  (PKIP) of Chinese Academy of Sciences (Grant No.~KJCX2.YW.W10).
\end{acknowledgments}

\bibliographystyle{apsrev4-1}
\bibliography{dark_energy,grb,misc}

\end{document}